# Meta-Programmable Analog Differentiator

Jérôme Sol[1], David R. Smith[2], and Philipp del Hougne[3*]

[1] INSA Rennes, CNRS, IETR - UMR 6164, F-35000, Rennes, France

[2] Center for Metamaterials and Integrated Plasmonics, Department of Electrical and Computer Engineering, Duke University, Durham, NC, 27708 USA

[3] Univ Rennes, CNRS, IETR - UMR 6164, F-35000, Rennes, France

* Correspondance to philipp.del-hougne@univ-rennes1.fr.

## Abstract

We show experimentally that the fundamental ingredient of wave-based signal differentiation, namely zeros of the scattering matrix that lie exactly on the real axis, can be imposed at will and *in situ* by purposefully perturbing an overmoded random scattering system. The resulting unprecedented flexibility overcomes current limitations of wave-based differentiators, both regarding their extreme vulnerability due to fabrication inaccuracies or environmental perturbations as well as their lack of *in situ* adaptability. Moreover, in addition to current miniaturization efforts, we suggest that integrability of wave processors can also be achieved by endowing existing bulky everyday-life systems that naturally scatter waves with a second signal-processing functionality. We demonstrate our technique by placing a programmable metasurface inside a 3D disordered metallic box: the hundreds of available degrees of freedom allow us to impose at will reflection zeros on a connected port, such that the reflected signal envelope is the temporal derivative of the incident one. We demonstrate our ability to toggle between differentiation of envelopes modulated onto distinct carriers. We also parallelize multiple differentiation operations on the same device and implement higher-order differentiators. Our "over-the-air" temporal differentiator for microwave carriers may find civilian and military applications in processing wireless communication or radar signals, for data segmentation and compression, as well as machine vision and hearing. Our generic concept is also applicable to optical, acoustic, and elastic scattering.

## Introduction

Differentiation is a pivotal mathematical operation with signal processing applications including edge-based segmentation for data compression or image sharpening, FM-to-AM demodulation for communication or Doppler-radar processing, as well as machine vision and hearing. Analog implementations of differentiation on wave processors promise much higher speeds and minimal power consumption in contrast to their digital electronic counterparts[1,2]. Yet, current wave-based differentiators are plagued by excessive sensitivity to fabrication inaccuracies and environmental conditions, and lack *in situ* adaptability. In this Article, instead of fabricating a carefully designed single-mode structure, we take an overmoded random scattering system as starting point and show that purposeful perturbations of its scattering properties, here with hundreds of degrees of freedom offered by an array of programmable meta-atoms[3,4], enable unprecedented flexibility and fidelity of "over-the-air" wave-based differentiators by imposing that zeros of the scattering matrix lie exactly on the real axis, at will and *in situ*.

Any linear wave system's input-output relation is a linear transformation that can be either a discretized or a continuous operator. The input and output channels can originate from various degrees of freedom, such as space (guided or propagating in free space), frequency and polarization, or even a mix thereof. The needs of artificial intelligence have recently driven the implementation of arbitrary matrix multiplications[5–11], mostly for discretized spatial channels[12]. At the same time, "basic" operations like differentiation continue to be of fundamental importance, for instance, in machine and human vision[13] for data compression via edge detection. These mathematical operations are typically implemented as continuous operators. A differentiator is a linear filter whose transfer function, $H(\omega) = i(\omega - \omega_0)$ for first-order differentiation of the envelope of a carrier at $\omega_0$, can be found in many natural and artificial special physical scattering scenarios. If information is encoded spatially such as in an optical image, then critical-coupling (CC) settings[14,15], the Brewster effect[16], and various (metamaterial) layer structures[17–20] have been shown to yield this transfer function. For temporally encoded information, implementations based on fiber gratings[21–24], critically coupled microring resonators[25], directional couplers[26], or interferometers[27–29] have been put forward. All these settings require very careful fabrication and/or alignment and are hence highly vulnerable to inaccuracies in the fabrication or operation environment. Wave-based differentiators are some of the most vulnerable wave processors because the underlying wave system is operated at a scattering anomaly ($H(\omega_0) = 0$); the origin of this extreme sensitivity can be traced back to a fundamental quantity in mesoscopic physics: the dwell time of the wave in the system[30]. We will elaborate on this link for the concrete case of CC for temporal differentiation in the following, which is also directly relevant to our experiments.

A longstanding question in wave physics and material science is how to excite a structure such that no wave energy is reflected back. CC is the simplest case of zero reflection and usually refers to coupling a single (guided) incident channel to an isolated mode of the structure[31,32], requiring perfect matching of the structure's excitation and decay rate. Moreover, perfect absorption has also been demonstrated for normal incidence of plane waves on thin metamaterials[33–35]. A generalized version of CC is coherent perfect absorption[36–40] (CPA) of multi-channel radiation by a resonant structure (with possibly overlapping modes). In all these cases, one zero of the scattering matrix is real-valued such that the structure can act as a steady-state sink: all incident radiation can be perfectly absorbed[1]. Since this zero's imaginary part vanishes, the dwell time of the corresponding wavefront in the system diverges[36,45–47].[2] The latter makes for extremely sensitive detectors[46,48], but conversely means that minute detuning of any system parameter moves the zero away from the real axis. As soon as the zero leaves the real axis, the spectrum's *linear* "V" shape and the associated *abrupt* $\pi$ phase jump in the vicinity of $\omega_0$ are no longer possible. Consequently, minute imperfections severely deteriorate and rapidly undermine the faithfulness of an analog differentiator, to the point that it becomes unsatisfactory, as illustrated in detail in Supplementary Note 2. This clarifies why tiny fabrication or alignment imperfections, or minute environmental perturbations, can severely impact the fidelity of a wave-based differentiator. Similarly, adapting, for instance, the filter's $\omega_0$ *in situ* in order to consecutively process envelopes on different carriers is generally very difficult or impossible. Wave processors realizing other functionalities which do not involve diverging dwell times are expected to display significantly less sensitivity.

---

[1] Because CPA has been introduced as the time-reverse of lasing (at threshold)[36], the loss mechanisms should originate from matter susceptibility rather than from undetected elastic scattering. Recently, there has been a growing interest in reflectionless coupling into (often quasi-lossless) structures which let radiation escape through unobserved channels[41–43]. Sometimes this has also been referred to as CPA[44] even though in these cases wave energy is *not* irreversibly transduced into other degrees of freedom such as heat. In the experiments we present, losses originate from absorption at the cavity boundaries (in line with the original definition of CPA), but the loss mechanism's nature is irrelevant for our general idea.

To overcome the excessive sensitivity of wave processors, a recent trend has been to explore ideas from topological photonics[49–51] but the lack of *in situ* adaptability remains. In parallel, the switching between different functionalities (differentiation, equation solving, etc.) has been studied[29,52,53] but this does not offer *in situ* adaptability of a given functionality, e.g. to the signal carrier, nor robustness to fabrication inaccuracies or environmental perturbations. A clear route to address the aforementioned challenges are wave processors that can be reprogrammed *in situ*. A notable application thereof to temporal differentiation involved a photonic integrated interferometer equipped with programmable phase modulators[29]. On paper, it appears that a single degree of freedom is sufficient to introduce a relative phase shift of $(2m + 1)\pi$ at $\omega_0$ between the two interferometer arms[29], yielding a perfect differentiator. In practice, however, the transfer function's magnitude minimum in Ref.[29] appears to be around 0.16; this system has a zero *close to* rather than *on* the real frequency axis, which jeopardizes a *linear* "V" shape and the associated *abrupt* π phase jump. Such limitations are inherent in the use of a single or few degrees of freedom in systems with a single or few resonances, and these limitations are also found in tunable microwave notch filters[54–57]. [3] The limitations originate from unaccounted non-idealities in real-life implementations such as coupling dispersion, transmission-line length variation, parasitic coupling, and other properties of the microstrip lines. Using a single or few degrees of freedom, it is certainly possible to tune few-resonance systems such that the zeros move in the complex plane; however, only under special conditions of $\mathcal{PT}$-symmetry[41,42] that are certainly not met by simple tunable notch filters (the uncompensated presence of absorption already trivially breaks $\mathcal{PT}$-symmetry) there is a guarantee that the zeros move exactly *on* the real frequency axis upon tuning. Hence, upon tuning simple notch filters in practice, zeros do not remain on the real frequency axis, and a few degrees of freedom tend to be insufficient to prevent that they drift away from the real frequency axis.

---

[2] The divergence of the dwell time can also be understood from considering the *phase delay time* which is defined as the frequency derivative of the scattering phase: $\tau = \mathrm{d}(\arg(S_{11}))/\mathrm{d}\omega$.[30] Given the discontinuity of the phase of the transfer function $i(\omega - \omega_0)$ at $\omega = \omega_0$, namely its phase jump of $\pi$, its derivative at $\omega = \omega_0$ diverges.

In contrast, our approach offers at least two orders of magnitude more degrees of freedom. This massive increase in programmability, together with the high density of zeros inherent to overmoded random scattering systems, makes it easy for us to perturb the system such that one of its zeros is placed, at a desired frequency, *on* the real frequency axis with extremely high precision (notch depth $< -70$ dB). Moreover, we can also switch to different functionalities and *simultaneously* create multiple zeros at *arbitrary* frequencies which is of importance for parallel wave processing at distinct frequencies, exploiting the wave equation's linearity[58]. Unlike conventional electronic processors, a single wave processor can simultaneously process various streams of information encoded on independent (spectral, polarization, etc.) channels[53,58–60], directly

---

[3] Tunable microwave notch filters are developed for applications other than analog differentiation where different metrics matter, resulting in filter shapes that are typically not only insufficiently deep but also oftentimes deformed with respect to the shape needed for analog differentiation.

multiplying its effective speed by the number of independent channels.

Our approach to rely on potentially bulky complex scattering systems such as a 3D disordered metallic box for microwave carriers may, at first sight, appear to be at odds with substantial efforts from the metamaterials community to miniaturize wave processors[17,61–64]. Indeed, the bulkiness of early optical processors, which relied on free-space propagation[65,66], but also the need for additional refractive elements (prisms, lenses) of many recent "flat" designs, thwarts integrability. Our technique, however, does not have any special requirements regarding the complex scattering system such that it can be implemented based on any *existing* bulky system whose primary functionality is not related to signal processing. We envision that ubiquitous metallic enclosures, such as a military toolbox or a microwave oven, can be endowed with a second signal processing functionality simply by inserting an ultrathin programmable metasurface[3,4] at an arbitrary location in order to tune their scattering properties. Thereby, we introduce a new perspective on integrability that decouples it from device volume and related miniaturization efforts. As such, our approach is arguably at least as convenient regarding integrability as are miniaturized wave processors. Its strength becomes particularly apparent for processing long-wavelength signals such as microwaves or sound in their native analog domain. In such scenarios relevant to radar, wireless communication, gesture recognition, ambient-assisted living, untethered virtual reality or voice-commanded devices, even metamaterial-based processors are cumbersome whereas our technique only requires the user to place an ultrathin programmable metasurface[3,4] at an arbitrary location inside an existing enclosure. A related alternative view on the integrability of wave processors was introduced in Ref.[9], but for less vulnerable spatially discrete monochromatic arbitrary matrix multiplications. Moreover, in Ref.[9] the configuration of the programmable metasurface was interpreted as input, that is, the wave processor did not operate in the signal's native domain, and furthermore averaging over multiple realizations was necessary as a consequence. If specific applications require far more compact implementations, it is also possible to implement our technique in flat

quasi-2D programmable chaotic cavities with currently available technology[67].

The technique underlying the present work to purposefully perturb a complex scattering system to impose real-valued scattering zeros "on demand" is, on the one hand, *conceptually* a topic of significant contemporary interest in mesoscopic physics. The concept was proposed and experimentally demonstrated in Ref.[48] for single-channel CPA in a lossy chaotic microwave cavity, with applications to precision sensing and secure communication. The idea was subsequently generalized to multiple channels[46,68], including cases with additional constraints on the allowed input wavefront[69]. The notion of parameter tuning is also central to various other recent works on scattering anomalies[42,70]. On the other hand, our specific *experimental* implementation in the microwave domain leveraging programmable metasurfaces relates to a large body of literature: these arrays of meta-atoms with individually reconfigurable scattering properties (usually reflection coefficient) are primarily used for free-space applications such as adaptive beamforming[3], holography[71], diffuse scattering[72], wireless communication[73–75], (intelligent) imaging[76–80,67,81,82] and spatio-temporal wave control[83]. However, they also find increasingly use inside rich scattering environments[84] as evidenced by various experiments on focusing[85–88], (sub-wavelength) sensing[89,90] and transmission matrix engineering[9,91–93]. Nonetheless, the generality of the discussed wave concepts implies that our idea can also be implemented in tunable acoustic or optical scattering systems[94–97].

In this paper, we establish the unprecedented flexibility offered for wave-based differentiators based on judiciously tuned overmoded random scattering systems, through a series of microwave experiments in a prototypical metallic disordered box equipped with 1-bit programmable metasurfaces. Our work is realized in the WLAN 5-GHz band and hence of potentially direct technological relevance. First, we demonstrate our ability to impose *in situ* a scattering zero at any desired frequency in this band, and to toggle at will between them. We directly inject various complicated waveforms in time-domain experiments to provide direct experimental evidence of our ability to compute the temporal derivative of an envelope of an arbitrary carrier (within the 5-GHz band). Second, we generalize this concept to toggling between multiple simultaneous zeros of the scattering matrix—that lie on the real axis—of various spectral separations in order to demonstrate parallel wave processing on the same device. Finally, we cascade two tunable chaotic cavities to compute second order derivatives, thereby engineering a transfer function usually associated with CPA exceptional points[98–101] that may also be enticing for applications in broadband near-perfect absorption.

## Results

**Operation principle.** To start, our goal is to compute the temporal derivative of a function $e(t)$ that modulates an arbitrary carrier $\omega_0$ by letting it bounce off the interface between a port and an irregularly shaped electrically large metallic box – see Figure 1. For the differentiation operation to happen, the system's transfer function $H(\omega)$ (i.e., the port's reflection spectrum) should approximate $i(\omega - \omega_0)$ in the vicinity of $\omega_0$, as sketched in the inset. A zero of the scattering matrix that lies on the real axis at $\omega_0$, which we define as a real-valued zero, yields this functional form of the reflection spectrum upon excitation[46,99]. However, the chances of observing a real-valued zero at the desired frequency $\omega_0$ within a random scattering system are extremely low[48]. By tuning a continuous system parameter, it is possible to observe a real-valued zero but most likely not at the targeted operating frequency[42]. Yet, our goal is to be able to re-program our analog wave-based differentiator *in situ* depending on the use case; specifically, as shown in Figure 1, we would like to adapt its transfer function to the incident carrier. To that end, we must be able to perfectly place a zero on the real-frequency axis "on demand" at any desired frequency.

To accomplish this goal, we mount two programmable metasurfaces[3,4] on the walls of the disordered box at arbitrary locations. These metasurfaces are ultrathin arrays of meta-atoms whose scattering properties can be controlled via a simple bias voltage. Our generic technique is not limited to implementations based on a specific programmable metasurface design, all that matters is that the programmable meta-atoms control as many rays as possible. Each meta-atom of the utilized prototype (see Supplementary Note 3 for technical details) has two possible states and is capable of roughly mimicking Dirichlet or Neuman boundary conditions under normal incidence[86]. By judiciously choosing the coding matrix of the programmable metasurface (see Supplementary Note 4), it is possible to impose a real-valued zero "on demand" at a desired frequency[48]. For instance, if the incident carrier is $\omega_1$ (blue), metasurface configuration $C_1$ (blue) is used; if the incident carrier is $\omega_2$ (red), metasurface configuration $C_2$ (red) is used, etc.

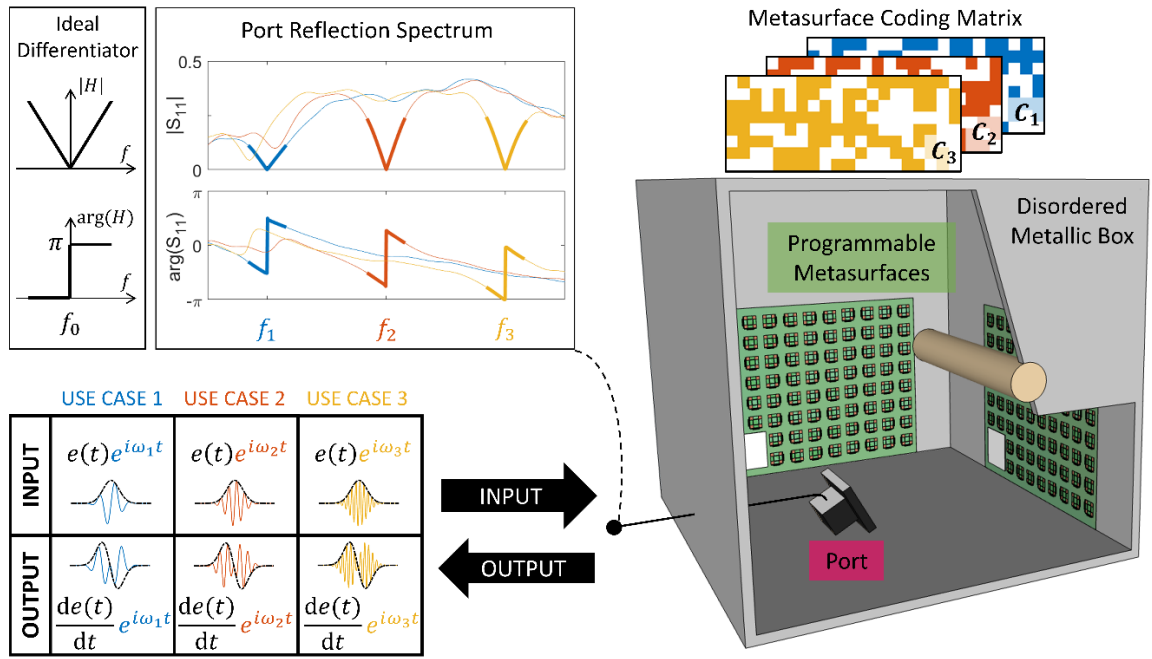

**Figure 1. Operation principle of the meta-programmable analog temporal differentiator.** A signal $e(t)$ is modulated as envelope onto a carrier $\omega_0$ and incident via a guided single-mode channel on a disordered metallic electrically large box. The scattering properties of the latter can be tuned via programmable metasurfaces mounted on its walls such that the port's reflection spectrum has a zero at the carrier frequency. Then, the port's transfer function matches that of an ideal differentiator, displaying in the zero's vicinity a *linear* "V" shape centered on $\omega_0$ as well as an abrupt

$\pi$ phase jump at $\omega_0$ (see inset). Consequently, the reflected signal's envelope is the temporal derivative of $e(t)$. Incident and reflected signals are separated via a circulator, see Supplementary Note 4 for details. A computer program digitally controls the realized differentiation operation by toggling between different metasurface configurations (color-coded) depending on the current needs in terms of the incident signal's carrier frequency.

**Direct observation of meta-programmable differentiation.** We begin by identifying eight metasurface configurations that yield a real-valued scattering matrix zero at eight distinct, evenly spaced frequencies between 5.05 GHz and 5.4 GHz. The resulting transfer functions are shown in Figure 2a,b. The desired linear "V" shapes, accompanied by abrupt phase shifts of $\pi$, are observed exactly at the eight chosen frequencies. In each case, the magnitude of the transfer function is below $-70$ dB at the central frequency (see Supplementary Note 5). The bandwidth over which the transfer functions are a good approximation to an ideal differentiator is around 15 MHz (see Supplementary Note 6). A global phase drift is observed and distinct for each operating frequency – however, as explained in Supplementary Note 1, this does not impact the desired differentiator functionality. Relative to the operating frequency, our fractional bandwidth of $3 \times 10^{-3}$ outperforms a number of previously reported temporal differentiators; only Ref.[22], a static non-programmable differentiator, achieves a considerably larger fractional bandwidth (see Supplementary Note 6). In comparison to the previously reported reconfigurable temporal differentiator from Ref.[29], our device's fractional bandwidth is an order of magnitude larger. More importantly, our device implements true real-valued zeros as evidenced by the depth of the reflection dips, their magnitude's linearity in the vicinity of the central frequency and the abruptness of the phase jump.

A direct experimental observation of meta-programmable differentiation requires the injection of various waveforms modulated onto various carriers and the observation of their reflections. We modulate each of the considered carriers in turn with one of three envelope functions shown in Figure 2c,f,i. For a given carrier, we toggle the metasurface to its corresponding configuration, we let the signal impinge on the port-cavity interface, and we measure the reflected signal – technical details are provided in Supplementary Notes 3 and 4. The first waveform, a Gaussian pulse (0.1 μs duration, 10 MHz bandwidth), is a typical function to test the quality of analog differentiators. Its derivative, if correctly computed, should be perfectly symmetric and have exactly zero amplitude at its center (see Figure 2d). In Figure 2e we superpose the envelopes of the output signals measured in the eight different carrier use cases. It is apparent that they are all highly similar and extremely close to the ideal analytical derivative. We also test our meta-programmable differentiator's performance with two more complicated functions: a quadratic polynomial and the skyline of Rennes, France. For these two functions, some less dominant spectral components lie outside the differentiator's operating bandwidth. In the former case (Figure 2h), the measured output signals faithfully reproduce the linear slopes expected for the derivative of a quadratic function, as well as the expected amplitudes. In the latter case (Figure 2k), the agreement with the analytical result is also very good; only for some very sharp peaks the measured magnitude is slightly below the expected one. Besides demonstrating differentiation based on a random scattering system, the major technological relevance of our work is the unprecedented flexibility of our differentiator: in this section, we leveraged its flexibility to toggle between differentiation for different carriers; in the following sections, we will further leverage this flexibility for more advanced and unique features.

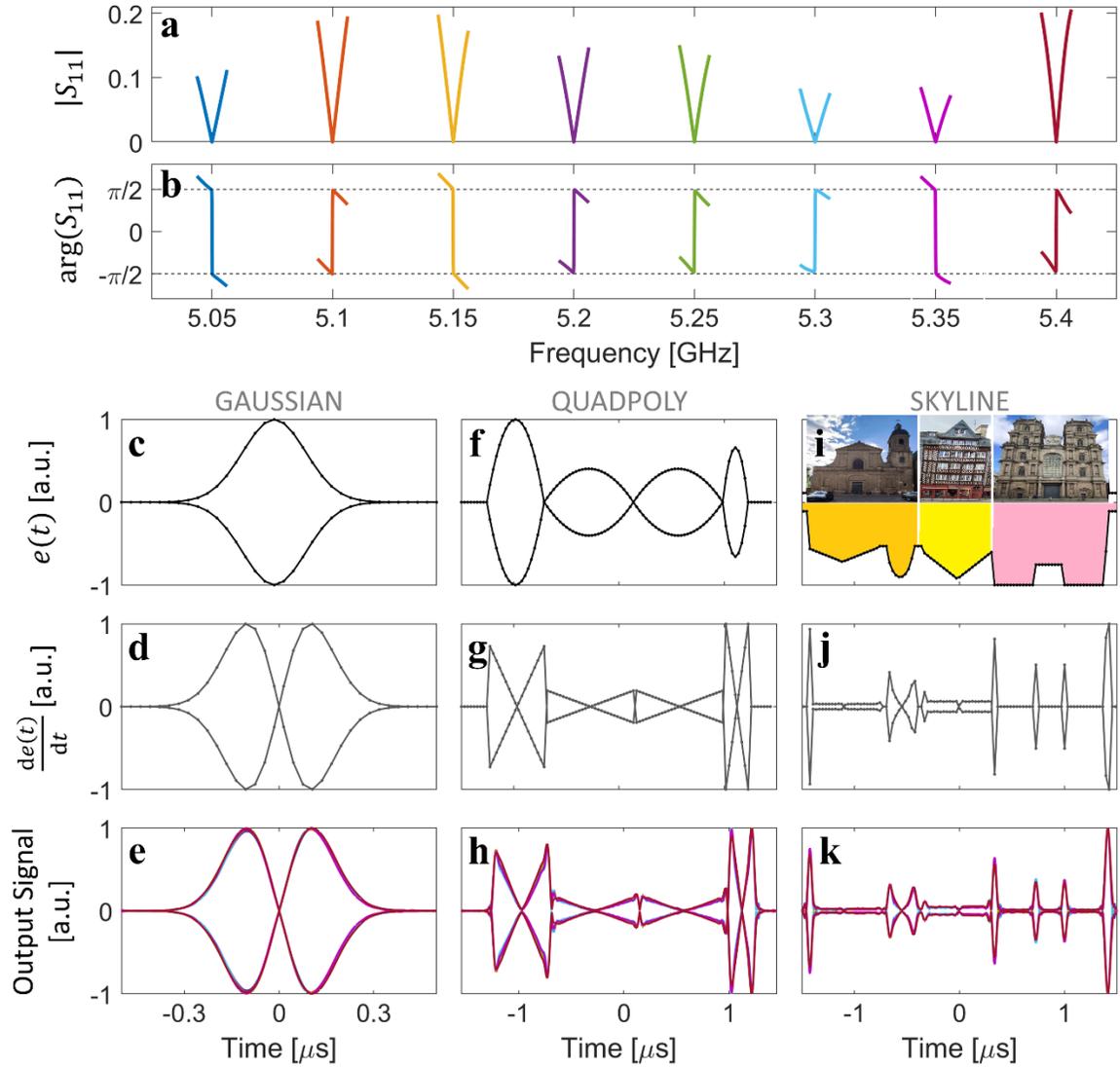

**Figure 2. Direct experimental results of meta-programmable analog temporal differentiation. a,b,** Amplitude (**a**) and phase (**b**) of the system's transfer function for eight different metasurface configurations (color-coded) that correspond to reflection zeros at eight equally spaced carrier frequencies in the 5-GHz band. **c,f,i,** Envelopes $e(t)$ of experimentally injected waveforms corresponding to a Gaussian function (**c**), a set of quadratic polynomial functions (**f**), and a skyline of Rennes, France, composed of Basilique Saint-Sauveur, a traditional timber-framed house and Cathédrale Saint-Pierre (**i**). **d,g,j,** Corresponding analytical derivatives $\frac{de(t)}{dt}$. **e,h,k,** Corresponding experimentally measured output signals. For a given waveform and carrier, the metasurface is toggled to the configuration optimized for this carrier, the input signal is injected, and the reflected signal is measured. The envelopes of the measured output signals for all eight considered

carriers are superposed on these figures using the same colors to identify different carriers as in (**a,b**).

**Parallelized meta-programmable differentiation.** A unique but to date largely underexploited advantage of wave processors over their electronic counterparts is that a single device can simultaneously process multiple data streams thanks to the linearity of the wave equation[53,58–60]. Thereby, its computational efficiency in terms of equivalent digital operations per unit time can be multiplied by the number of independent data streams. The differentiation operation that we are concerned with in this paper can be parallelized, for instance, by encoding different data streams onto different carriers. The reflection spectrum of the system should then *simultaneously* present *multiple* real-valued scattering zeros corresponding to the targeted carrier frequencies. The experimental setup remains that from Figure 1, except that the impinging wave is now the sum of multiple independent carriers, each modulated by an independent data stream. This principle is illustrated in Figure 3a. Our *parallelized* meta-programmable analog differentiator can toggle between different simultaneously targeted carrier frequencies. As shown in Figure 3a, for a first use case, a blue metasurface configuration may be used that imposes simultaneous real-valued scattering zeros for carriers $\omega_{1,A}$ and $\omega_{1,B}$; for a second use case, a red metasurface configuration may be used for carriers $\omega_{2,A}$ and $\omega_{2,B}$, etc.

In Figure 3b,c,f,g,j,k, three examples of experimental measurements of optimized transfer functions with two simultaneous real-valued zeros at different frequencies are shown. In all cases, the transfer function magnitude displays the linear "V" shape and the phase the corresponding abrupt $\pi$ jump. Examples with more than two simultaneous zeros are shown in Supplementary Note 8 but in the following we focus on two simultaneous zeros to directly test the meta-programmable parallelized differentiation ability of our system. To that end, we generate two independent signals and inject their sum into our system which is toggled to a state that corresponds to the two chosen carriers. We measure the reflected signal and bandpass-filter it around each of the two carriers, respectively. The resulting two demultiplexed data streams are shown in Figure

3d,e,h,i,l,m. The obtained signal output envelops match very well the analytically expected ones for both carriers in all three cases. These results illustrate our ability to perform parallelized meta-programmable differentiation. Ref.[29] based on a tunable interferometer did not discuss parallel computing but it is clear that if at all possible, the simultaneous real-valued zeros would have to be spaced according to the free spectral range as opposed to being able to place them at arbitrary locations as in our system. If the carrier frequencies' separation exceeds the metasurface's operation bandwidth, separate metasurfaces (one for each carrier) may be deployed. In that case, only the designated metasurface would modulate the corresponding carrier frequency such that the different metasurface configurations could be optimized independently from each other[9].

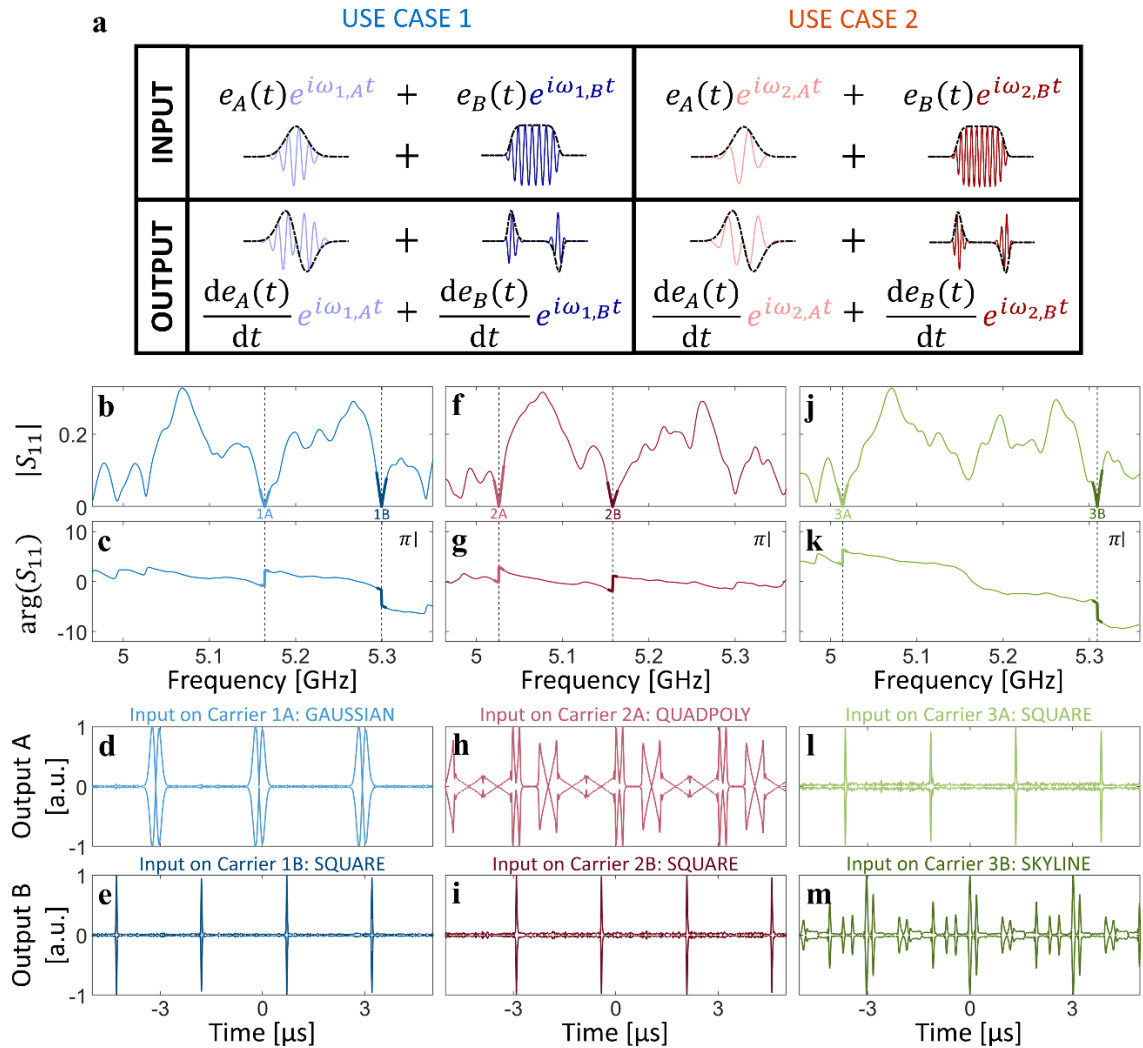

**Figure 3. Parallelization of meta-programmable wave-based differentiation. a,** Principle of parallel computing with spectral degrees of freedom. The injected waveform is the sum of two signals (two independent envelopes $e_A(t)$ and $e_B(t)$ modulated onto distinct carriers $\omega_{1,A}$ and $\omega_{1,B}$) and the reflected signal is the sum of the derivatives of these two envelopes, modulated onto the respective carriers. **b,c,f,g,j,k,** Amplitude and phase of the system's transfer function for three examples (color-coded) of choices of two *simultaneously* imposed reflection zeros. The computer program can toggle between these by changing the metasurface configuration. **d,e,h,i,l,m,** Envelope of experimentally measured output signal (spectrally bandpass-filtered around the respective carrier frequencies) upon injecting the indicated waveforms (see Supplementary Note 4 for details) in the three considered use cases.

**Higher-order meta-programmable differentiation.** To this point, we have considered only first-order temporal differentiation; in this section, we generalize the previous concepts to higher-order differentiation. An $n$-th-order differentiator's transfer function is $[i(\omega - \omega_0)]^n$, which can be physically implemented in various manners. For instance, instead of tuning our system to CPA, one could attempt to tune it to a CPA exceptional point (EP) at which two eigenvalues *and* eigenvectors of the wave operator associated with purely incoming boundary conditions coalesce[99]; such a CPA-EP would yield the transfer function associated with a second-order differentiator (see inset in Figure 4). However, preliminary tests following this route were unsuccessful, indicating that more degrees of freedom and/or more elaborate optimization protocols may be necessary. Another route to obtain the desired transfer function of a second-order differentiator is to cascade two first-order differentiators. If the input signal was known to have a sufficient temporal sparsity, a combination of directional couplers and delay loop could be added to the setup in Figure 1 to realize the higher-order differentiation via multiple passes across the same port-cavity interface. A more generic implementation is to physically cascade multiple setups akin to the one from Figure 1, all tuned to simultaneously have a real-valued scattering zero at the same frequency, as shown in

Figure 4a. We adopt the latter approach in the following.

Our goal is now to simultaneously impose a real-valued scattering zero at the same frequency in two similar but distinct chaotic cavities, and to toggle between different examples thereof that differ in terms of the chosen frequency, in order to then directly illustrate our ability to perform meta-programmable second-order differentiation. Given that our experimentally available number of programmable meta-atoms is limited, we now split them equally between the two cavities. The optimization problem is now much harder since we have only half of the previously available degrees of freedom in each cavity. This is of course not a general limitation of our concept but related to our experimental constraints, and we cater to it by allowing some flexibility regarding the chosen frequency in our optimization protocol. In the left columns of Figures 4b,c,d,e, we report four examples of magnitude and phase of experimentally measured transfer functions for second-order differentiation. The flat phase (modulo $2\pi$ and ignoring the irrelevant background phase drift) matches that of the ideal transfer function shown in the inset of Figure 4a. The magnitudes also display the desired quadratic behavior in the vicinity of $\omega_0$ but appear to offer a more limited useable bandwidth than those from Figure 2a. We directly test our meta-programmable second-order differentiator by injecting the Gaussian signal from Figure 2c, modulated onto different carriers, and in each case toggling the metasurfaces to the configuration corresponding to the carrier. Compared to first-order differentiation, the output signals are much weaker and hence noisier, such that we average the output measurements over 20 acquisitions. The averaged output envelopes for the four considered carriers in Figure 4b,c,d,e are close to the analytically expected output (see Figure 4a). The symmetry with respect to the central peak is very good, and the two dips on either side go as low as possible given the measurement noise. The measurement noise is also evident far away from the central peak where the output signal is analytically zero.

Beyond the wave computing functionality discussed so far, the setup from Figure 4 emulates, as previously noted, the transfer function of CPA-EPs which are thus far mainly investigated for broadband near-perfect absorption applications. In contrast to the intensity-dependent nonlinear CPA-EP in Ref.[100], our experiment offers this transfer function irrespective of the incident power signal. Moreover, the homogeneous absorption in our system due to Ohmic losses on the walls, as opposed to localized absorption mechanisms, enables the absorption of high-power incident signals without significant localized heating. Furthermore, the programmability of our setup regarding the CPA-EP frequency may be important for future absorption applications, too.

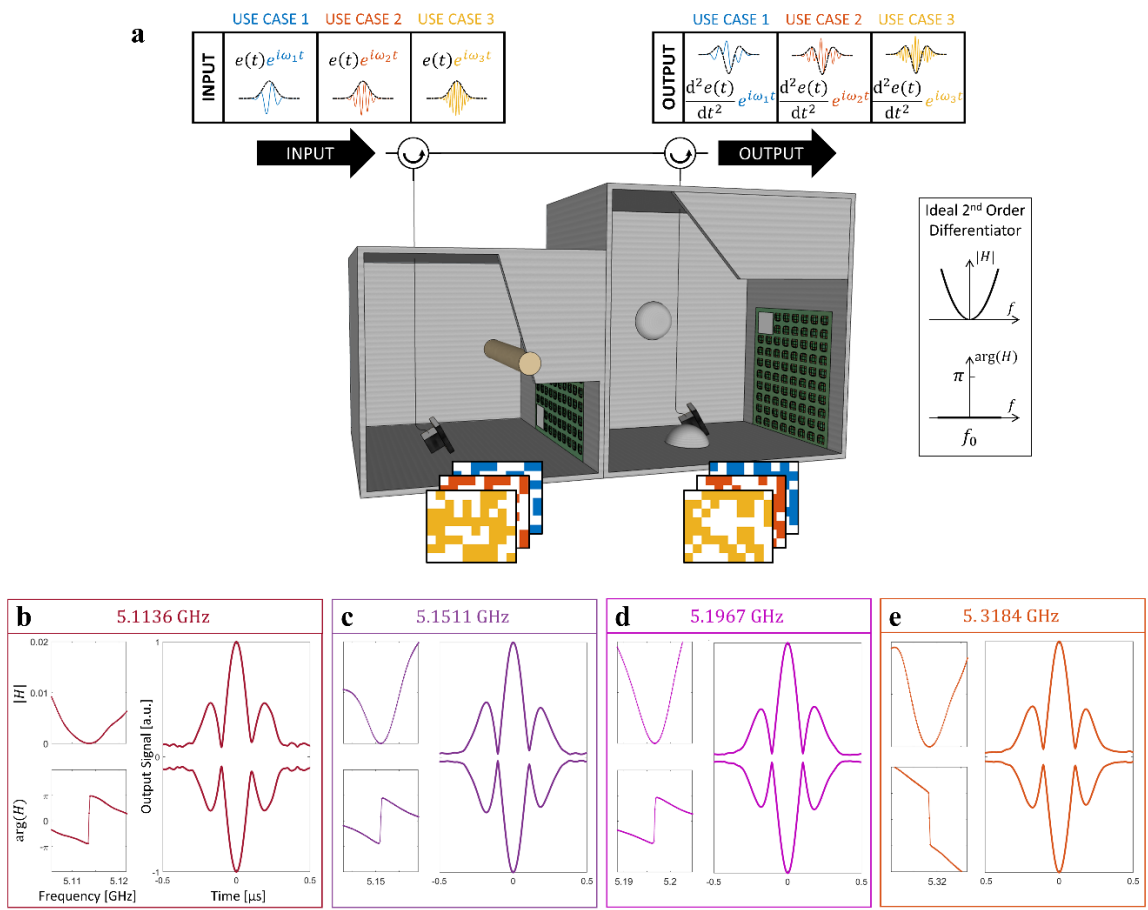

**Figure 4. Meta-programmable second-order differentiator. a,** Operation principle. Two setups akin to the one from Figure 1 are cascaded (using circulators as shown) in order to implement the transfer function associated with an ideal second-order differentiator (see inset). Again, the computer program can toggle between metasurface configurations such that second-order derivatives of signal envelopes are computed for different carriers (color-coded). **b-e,** Examples of four

experimentally measured use cases corresponding to the four distinct indicated carrier frequencies. In each case, amplitude and phase of the system's transfer function are shown, as well as the experimentally measured output signal envelope upon injection of a Gaussian pulse. The displayed output signal envelopes are averaged over 20 acquisitions to alleviate the impact of measurement noise.

## Discussion

Besides speed and the ease of processing signals in their native domain, a compelling argument for wave-based computing is its energy efficiency. Indeed, a passive material can offer a desired transfer function without any energy consumption. However, a passive material cannot serve as the basis of a fair comparison with our technique since a passive material lacks all the features and advantages of reconfigurability that motivate our present work. Our device is "quasi-passive": the programmable metasurface does *not* provide any energy to the wave and only requires minimal power, as low as a few µW per meta-atom[85], to maintain the meta-atom in the desired state or to flip it. Future metasurface technologies, such as those based on chalcogenide glasses, may even overcome the need for energy to maintain a state and only require minimal energy to flip the state. All this is in contrast to the proposed reconfigurable photonic interferometer-based processor from Ref.[29] which, besides phase modulators, also intrinsically relies on a multitude of signal amplifiers.

To summarize, we demonstrated experimentally the essential steps to implement a meta-programmable "over-the-air" wave-based differentiator. The unprecedented flexibility of our approach is enabled by not fabricating a carefully designed device with a single resonance and a few degrees of freedom but leveraging *in situ* the massive amount of degrees of freedom offered by a programmable metasurface inside a complex scattering system in order to tune scattering zeros onto the real axis at will. The resulting *in situ* adaptability overcomes the extreme sensitivity of analog differentiators to any sort of detuning which originates from the fact that they operate at a scattering anomaly associated with a diverging dwell time. We demonstrated with direct temporal measurements of the output signal that our technique can (i) toggle between operation at different carrier frequencies, (ii) parallelize the computational operation based on the wave equation's linearity, and (iii) implement higher-order differentiation. Our approach can be extended to process spatially encoded information, to using overmoded random scattering systems based on *guided* waves, and to optical or acoustic scattering, as detailed in Supplementary Note 9. Besides the implementation of analog differentiation, the unprecedented flexibility and precision of the implemented filters is relevant to applications in wideband defense systems such as cognitive radio.

Looking forward, we envision that a frequency sensing module is integrated with the entire system such that the latter can autonomously adapt itself to the incident signal; such an "intelligent" control of the metasurface configuration based on sensor measurements is easily implemented with current technology[102,103]. We also note that the switching time between metasurface configurations can be as low as 20 µs with state-of-the-art microcontrollers (see, for instance, Ref.[74]). The proposed meta-programmable differentiator can also serve as the pivotal ingredient of analog solvers of differential equations[104].

Another avenue that can prove fruitful is to explore potential benefits of continuously programmable meta-atoms[77] as well as multi-channel CPA[46,68] for meta-programmable analog differentiators. The former may drastically simplify gradient-based search methods for the optimal configuration, and hence be worthwhile the additional electronic burden. The latter may, according to indications in our ongoing work, facilitate imposing real-valued zeros in our system which potentially offsets the requirement for multi-channel amplitude and phase control, or an additional constraint to make the CPA eigenvector coincide with a predefined one[69].

# Methods

**Experimental setup.** A photographic image of the experimental setup from Figure 1 is shown in Supplementary Figure 2. A waveguide-to-coax adapter (RA13PBZ012-B-SMA-F) couples a guided wave to an electrically large disordered metallic box. Two programmable metasurfaces, each containing 76 meta-atoms, cover 16.2% of the cavity surface. Within a 400 MHz interval centered on 5.2 GHz, this metasurface can efficiently manipulate the scattered field. Each meta-atom offers independent 1-bit control over two orthogonal field polarizations, mimicking approximately Dirichlet or Neuman boundary conditions under normal incidence. Ohmic losses on the cavity walls result in a quality factor of the enclosure on the order of 410 such that approximately 21 modes overlap at a given frequency on average. Further details about the experimental setup and the characterization of the metasurface prototype are provided in Supplementary Note 4. Note that our concept is generic and could be implemented with any other programmable metasurface design, too.

**Determination of metasurface configurations.** Identifying a metasurface configuration that yields a desired scattering response, e.g., a real-valued scattering zero at a desired frequency, is an inverse problem. For our proof-of-principle experiments, we solved this inverse problem via an iterative experimental trial-and-error algorithm detailed in Supplementary Note 4 which involved roughly 800 measurements with a vector network analyzer (Agilent Technologies PNA-L Network Analyzer N5230C, 0 dBm emitted power, 10 kHz intermediate-frequency bandwidth). In the future, the inverse design of metasurface configurations can be performed much faster and in software, once an artificial neural network has been trained to approximate the function that maps the configuration to the scattering response. Ref.[80] already employs a learned forward model to predict the scattering response of a programmable metasurface, albeit in quasi free space rather than inside a complex scattering enclosure. Next-generation meta-atoms with fine-grained programmability (> 1-bit) will expedite the optimization through their compatibility with continuous gradient-descent protocols. In certain settings without environmental perturbations during runtime, the identification of suitable metasurface configurations can be completed offline during a calibration phase and presents no burden during runtime. Generally speaking, when the optimization objectives are challenging and the available number of degrees of freedom is limited, in many applications one can find significant constraint relaxations such as not requiring the zero to occur at a specific frequency but within a certain interval.

**Time-domain experiments.** For a given test scenario involving a specific function $e(t)$ to be derived and a specific carrier $\omega_0$, we generated the waveform $e(t)e^{i\omega_0 t}$ with a signal generator (Aeroflex IFR 3416, 2 dBm emitted power, 33 MHz sampling rate) and toggled the metasurface to the configuration suitable for the chosen carrier. The input signal was injected via Port 1 of a circulator (PE83CR006; see Supplementary Figure 4b), impinged on the system interface via Port 2 of the circulator, and the output signal which exited through Port 3 of the circulator was measured on an oscilloscope (SDA 816Zi-B 16GHz Serial Data Analyzer, 40GS/s sampling rate). For the experiments on parallelized computing, a second signal generator (Agilent MXG Analog N5183A) generated the second input signal, and the two input signals were summed before being injected via Port 1 of the circulator. More details on the time-domain experiments can be found in Supplementary Note 4.

## Data Availability

The data that support the findings of this study are available from the corresponding author upon request.

## Code Availability

Code that supports the findings of this study is available upon reasonable request from the corresponding author.

## Author Contributions

P.d.H. conceived the project. J.S. and P.d.H. conducted the experiments. D.R.S. and P.d.H. interpreted the results. All authors contributed with thorough discussions. P.d.H. wrote the manuscript.

## Acknowledgements

The authors thank Steven M. Anlage, Thomas M. Antonsen, Matthieu Davy, Edward Ott and A. Douglas Stone for stimulating discussions, and François Yven for helpful exchanges regarding the time-domain measurement setup. The authors acknowledge funding from the European Union through the European Regional Development Fund (ERDF), and the French region of Brittany and Rennes Metropole through the CPER Project SOPHIE/STIC & Ondes. The metasurface prototypes were purchased from Greenerwave.

## Additional Information

**Supplementary Information** accompanies this article.
**Competing Interests**: The authors declare no competing interests.

# References

1. Solli, D. R. & Jalali, B. Analog optical computing. *Nat. Photonics* **9**, 704–706 (2015).
2. Zangeneh-Nejad, F., Sounas, D. L., Alù, A. & Fleury, R. Analogue computing with metamaterials. *Nat. Rev. Mater.* **6**, 207–225 (2021).
3. Cui, T. J., Qi, M. Q., Wan, X., Zhao, J. & Cheng, Q. Coding metamaterials, digital metamaterials and programmable metamaterials. *Light Sci. Appl.* **3**, e218–e218 (2014).
4. Kaina, N., Dupré, M., Fink, M. & Lerosey, G. Hybridized resonances to design tunable binary phase metasurface unit cells. *Opt. Express* **22**, 18881 (2014).
5. Miller, D. A. B. Self-configuring universal linear optical component [Invited]. *Photonics Res.* **1**, 1 (2013).
6. Carolan, J. *et al.* Universal linear optics. *Science* **349**, 711–716 (2015).
7. Ribeiro, A., Ruocco, A., Vanacker, L. & Bogaerts, W. Demonstration of a 4 × 4-port universal linear circuit. *Optica* **3**, 1348 (2016).
8. Shen, Y. *et al.* Deep learning with coherent nanophotonic circuits. *Nat. Photonics* **11**, 441–446 (2017).
9. del Hougne, P. & Lerosey, G. Leveraging Chaos for Wave-Based Analog Computation: Demonstration with Indoor Wireless Communication Signals. *Phys. Rev. X* **8**, 041037 (2018).
10. Matthès, M. W., del Hougne, P., de Rosny, J., Lerosey, G. & Popoff, S. M. Optical complex media as universal reconfigurable linear operators. *Optica* **6**, 465 (2019).
11. Buddhiraju, S., Dutt, A., Minkov, M., Williamson, I. A. D. & Fan, S. Arbitrary linear transformations for photons in the frequency synthetic dimension. *Nat. Commun.* **12**, 2401 (2021).
12. Moeini, M. M. & Sounas, D. L. Discrete space optical signal processing. *Optica* **7**, 1325 (2020).
13. Hubel, D. H. *Eye, Brain, and Vision*. (Scientific American, 1995).
14. Zhu, T. *et al.* Plasmonic computing of spatial differentiation. *Nat. Commun.* **8**, 15391 (2017).
15. Wesemann, L. *et al.* Selective near-perfect absorbing mirror as a spatial frequency filter for optical image processing. *APL Photonics* **4**, 100801 (2019).
16. Youssefi, A., Zangeneh-Nejad, F., Abdollahramezani, S. & Khavasi, A. Analog computing by Brewster effect. *Opt. Lett.* **41**, 3467 (2016).
17. Silva, A. *et al.* Performing Mathematical Operations with Metamaterials. *Science* **343**, 160–163 (2014).
18. Doskolovich, L. L., Bykov, D. A., Bezus, E. A. & Soifer, V. A. Spatial differentiation of optical beams using phase-shifted Bragg grating. *Opt. Lett.* **39**, 1278 (2014).
19. Zangeneh-Nejad, F., Khavasi, A. & Rejaei, B. Analog optical computing by half-wavelength slabs. *Opt. Commun.* **407**, 338–343 (2018).
20. Zangeneh-Nejad, F. & Fleury, R. Performing mathematical operations using high-index acoustic metamaterials. *New J. Phys.* **20**, 073001 (2018).
21. Kulishov, M. & Azaña, J. Long-period fiber gratings as ultrafast optical differentiators. *Opt. Lett.* **30**, 2700 (2005).
22. Slavík, R., Park, Y., Kulishov, M., Morandotti, R. & Azaña, J. Ultrafast all-optical differentiators. *Opt. Express* **14**, 10699 (2006).
23. Berger, N. K. *et al.* Temporal differentiation of optical signals using a phase-shifted fiber Bragg

grating. *Opt. Express* **15**, 371 (2007).

24. Li, M., Janner, D., Yao, J. & Pruneri, V. Arbitrary-order all-fiber temporal differentiator based on a fiber Bragg grating: design and experimental demonstration. *Opt. Express* **17**, 19798 (2009).
25. Liu, F. *et al.* Compact optical temporal differentiator based on silicon microring resonator. *Opt. Express* **16**, 15880 (2008).
26. Huang, T. L., Zheng, A. L., Dong, J. J., Gao, D. S. & Zhang, X. L. Terahertz-bandwidth photonic temporal differentiator based on a silicon-on-isolator directional coupler. *Opt. Lett.* **40**, 5614 (2015).
27. Hsue, C.-W., Tsai, L.-C. & Chen, K.-L. Implementation of First-Order and Second-Order Microwave Differentiators. *IEEE Trans. Microwave Theory Techn.* **52**, 1443–1448 (2004).
28. Park, Y., Azaña, J. & Slavík, R. Ultrafast all-optical first- and higher-order differentiators based on interferometers. *Opt. Lett.* **32**, 710 (2007).
29. Liu, W. *et al.* A fully reconfigurable photonic integrated signal processor. *Nat. Photonics* **10**, 190–195 (2016).
30. Rotter, S. & Gigan, S. Light fields in complex media: Mesoscopic scattering meets wave control. *Rev. Mod. Phys.* **89**, 015005 (2017).
31. Cai, M., Painter, O. & Vahala, K. J. Observation of Critical Coupling in a Fiber Taper to a Silica-Microsphere Whispering-Gallery Mode System. *Phys. Rev. Lett.* **85**, 74–77 (2000).
32. Yariv, A. Critical coupling and its control in optical waveguide-ring resonator systems. *IEEE Photon. Technol. Lett.* **14**, 483–485 (2002).
33. Landy, N. I., Sajuyigbe, S., Mock, J. J., Smith, D. R. & Padilla, W. J. Perfect Metamaterial Absorber. *Phys. Rev. Lett.* **100**, 207402 (2008).
34. Watts, C. M., Liu, X. & Padilla, W. J. Metamaterial Electromagnetic Wave Absorbers. *Adv. Mater.* **24**, OP98–OP120 (2012).
35. Asadchy, V. S. *et al.* Broadband Reflectionless Metasheets: Frequency-Selective Transmission and Perfect Absorption. *Phys. Rev. X* **5**, 031005 (2015).
36. Chong, Y. D., Ge, L., Cao, H. & Stone, A. D. Coherent Perfect Absorbers: Time-Reversed Lasers. *Phys. Rev. Lett.* **105**, 053901 (2010).
37. Wan, W. *et al.* Time-reversed lasing and interferometric control of absorption. *Science* **331**, 889–892 (2011).
38. Baranov, D. G., Krasnok, A., Shegai, T., Alù, A. & Chong, Y. Coherent perfect absorbers: linear control of light with light. *Nat. Rev. Mater.* **2**, 17064 (2017).
39. Krasnok, A. *et al.* Anomalies in light scattering. *Adv. Opt. Photonics* **11**, 892 (2019).
40. Li, H., Suwunnarat, S., Fleischmann, R., Schanz, H. & Kottos, T. Random Matrix Theory Approach to Chaotic Coherent Perfect Absorbers. *Phys. Rev. Lett.* **118**, 044101 (2017).
41. Dhia, A.-S. B.-B., Chesnel, L. & Pagneux, V. Trapped modes and reflectionless modes as eigenfunctions of the same spectral problem. *Proc. R. Soc. A.* **474**, 20180050 (2018).
42. Sweeney, W. R., Hsu, C. W. & Stone, A. D. Theory of reflectionless scattering modes. *Phys. Rev. A* **102**, 063511 (2020).
43. Grimm, P., Razinskas, G., Huang, J.-S. & Hecht, B. Driving plasmonic nanoantennas at perfect impedance matching using generalized coherent perfect absorption. *Nanophotonics* **10**, 1879–1887 (2021).
44. Pichler, K. *et al.* Random anti-lasing through coherent perfect absorption in a disordered medium.

*Nature* **567**, 351–355 (2019).

45. Asano, M. *et al.* Anomalous time delays and quantum weak measurements in optical microresonators. *Nat. Commun.* **7**, 13488 (2016).
46. del Hougne, P., Yeo, K. B., Besnier, P. & Davy, M. On-Demand Coherent Perfect Absorption in Complex Scattering Systems: Time Delay Divergence and Enhanced Sensitivity to Perturbations. *Laser Photonics Rev.* **15**, 2000471 (2021).
47. Chen, L., Anlage, S. M. & Fyodorov, Y. V. Generalization of Wigner time delay to subunitary scattering systems. *Phys. Rev. E* **103**, L050203 (2021).
48. F. Imani, M., Smith, D. R. & del Hougne, P. Perfect Absorption in a Disordered Medium with Programmable Meta-Atom Inclusions. *Adv. Funct. Mater.* **30**, 2005310 (2020).
49. Zhu, T. *et al.* Topological optical differentiator. *Nat. Commun.* **12**, 680 (2021).
50. Zangeneh-Nejad, F. & Fleury, R. Topological analog signal processing. *Nat. Commun.* **10**, 2058 (2019).
51. Zangeneh-Nejad, F. & Fleury, R. Disorder-Induced Signal Filtering with Topological Metamaterials. *Adv. Mater.* **32**, 2001034 (2020).
52. Li, M. *et al.* Reconfigurable Optical Signal Processing Based on a Distributed Feedback Semiconductor Optical Amplifier. *Sci. Rep.* **6**, 19985 (2016).
53. Momeni, A., Rouhi, K. & Fleury, R. Switchable and Simultaneous Spatiotemporal Analog Computing. *arXiv:2104.10801 [physics]* (2021).
54. Gil, I. *et al.* Varactor-loaded split ring resonators for tunable notch filters at microwave frequencies. *Electron. Lett.* **40**, 1347 (2004).
55. Ou, Y.-C. & Rebeiz, G. M. Lumped-Element Fully Tunable Bandstop Filters for Cognitive Radio Applications. *IEEE Trans. Microwave Theory Techn.* **59**, 2461–2468 (2011).
56. Wu, Z., Shim, Y. & Rais-Zadeh, M. Miniaturized UWB Filters Integrated With Tunable Notch Filters Using a Silicon-Based Integrated Passive Device Technology. *IEEE Trans. Microwave Theory Techn.* **60**, 518–527 (2012).
57. Jeong, S.-W. & Lee, J. Frequency- and Bandwidth-Tunable Bandstop Filter Containing Variable Coupling Between Transmission Line and Resonator. *IEEE Trans. Microwave Theory Techn.* **66**, 943–953 (2018).
58. Camacho, M., Edwards, B. & Engheta, N. A single inverse-designed photonic structure that performs parallel computing. *Nat. Commun.* **12**, 1466 (2021).
59. Babaee, A., Momeni, A., Abdolali, A. & Fleury, R. Parallel Analog Computing Based on a 2 × 2 Multiple-Input Multiple-Output Metasurface Processor With Asymmetric Response. *Phys. Rev. Applied* **15**, 044015 (2021).
60. Zhou, Y. *et al.* Analogue Optical Spatiotemporal Differentiator. *Adv. Opt. Mater.* **9**, 2002088 (2021).
61. Kwon, H., Sounas, D., Cordaro, A., Polman, A. & Alù, A. Nonlocal Metasurfaces for Optical Signal Processing. *Phys. Rev. Lett.* **121**, 173004 (2018).
62. Mohammadi Estakhri, N., Edwards, B. & Engheta, N. Inverse-designed metastructures that solve equations. *Science* **363**, 1333–1338 (2019).
63. Cordaro, A. *et al.* High-Index Dielectric Metasurfaces Performing Mathematical Operations. *Nano Lett.* **19**, 8418–8423 (2019).
64. Zhou, Y., Zheng, H., Kravchenko, I. I. & Valentine, J. Flat optics for image differentiation. *Nat.*

*Photonics* **14**, 316–323 (2020).

65. Leger, J. R. & Lee, S. H. Coherent optical implementation of generalized two-dimensional transforms. *Opt. Eng.* **18**, 185518 (1979).
66. Reck, M., Zeilinger, A., Bernstein, H. J. & Bertani, P. Experimental realization of any discrete unitary operator. *Phys. Rev. Lett.* **73**, 58–61 (1994).
67. Sleasman, T. *et al.* Implementation and characterization of a two-dimensional printed circuit dynamic metasurface aperture for computational microwave imaging. *IEEE Trans. Antennas Propag.* **69**, 2151 (2020).
68. Frazier, B. W., Antonsen, T. M., Anlage, S. M. & Ott, E. Wavefront shaping with a tunable metasurface: Creating cold spots and coherent perfect absorption at arbitrary frequencies. *Phys. Rev. Research* **2**, 043422 (2020).
69. del Hougne, P., Yeo, K. B., Besnier, P. & Davy, M. Coherent Wave Control in Complex Media with Arbitrary Wavefronts. *Phys. Rev. Lett.* **126**, 193903 (2021).
70. Kang, Y. & Genack, A. Z. Transmission zeros with topological symmetry in complex systems. *Phys. Rev. B* **103**, L100201 (2021).
71. Li, L. *et al.* Electromagnetic reprogrammable coding-metasurface holograms. *Nat. Commun.* **8**, 197 (2017).
72. Moccia, M. *et al.* Coding Metasurfaces for Diffuse Scattering: Scaling Laws, Bounds, and Suboptimal Design. *Adv. Opt. Mater.* **5**, 1700455 (2017).
73. Tang, W., Wong, K.-K., Li, X., Zhao, X. & Jin, S. MIMO Transmission Through Reconfigurable Intelligent Surface: System Design, Analysis, and Implementation. *IEEE J. Sel. Areas Commun.* **38**, 17 (2020).
74. Zhao, H. *et al.* Metasurface-assisted massive backscatter wireless communication with commodity Wi-Fi signals. *Nat. Commun.* **11**, 3926 (2020).
75. Dai, L. *et al.* Reconfigurable Intelligent Surface-Based Wireless Communications: Antenna Design, Prototyping, and Experimental Results. *IEEE Access* **8**, 45913–45923 (2020).
76. Sleasman, T., F. Imani, M., Gollub, J. N. & Smith, D. R. Dynamic metamaterial aperture for microwave imaging. *Appl. Phys. Lett.* **107**, 204104 (2015).
77. Sleasman, T., Imani, M. F., Gollub, J. N. & Smith, D. R. Microwave Imaging Using a Disordered Cavity with a Dynamically Tunable Impedance Surface. *Phys. Rev. Applied* **6**, 054019 (2016).
78. Li, L. *et al.* Intelligent metasurface imager and recognizer. *Light Sci. Appl.* **8**, 97 (2019).
79. del Hougne, P., Imani, M. F., Diebold, A. V., Horstmeyer, R. & Smith, D. R. Learned Integrated Sensing Pipeline: Reconfigurable Metasurface Transceivers as Trainable Physical Layer in an Artificial Neural Network. *Adv. Sci.* **7**, 1901913 (2019).
80. Li, H.-Y. *et al.* Intelligent Electromagnetic Sensing with Learnable Data Acquisition and Processing. *Patterns* **1**, 100006 (2020).
81. Li, W., Qi, J. & Sihvola, A. Meta-Imaging: from Non-Computational to Computational. *Adv. Opt. Mater.* **8**, 2001000 (2020).
82. Saigre-Tardif, C., Faqiri, R., Zhao, H., Li, L. & del Hougne, P. Intelligent Meta-Imagers: From Compressed to Learned Sensing. *arXiv:2110.14022* (2021).
83. Zhang, L. *et al.* Space-time-coding digital metasurfaces. *Nat. Commun.* **9**, 4334 (2018).
84. Alexandropoulos, G. C., Shlezinger, N. & del Hougne, P. Reconfigurable Intelligent Surfaces for

Rich Scattering Wireless Communications: Recent Experiments, Challenges, and Opportunities. *IEEE Commun. Mag.* **59**, 28–34 (2021).

85. Kaina, N., Dupré, M., Lerosey, G. & Fink, M. Shaping complex microwave fields in reverberating media with binary tunable metasurfaces. *Sci. Rep.* **4**, 6693 (2015).
86. Dupré, M., del Hougne, P., Fink, M., Lemoult, F. & Lerosey, G. Wave-Field Shaping in Cavities: Waves Trapped in a Box with Controllable Boundaries. *Phys. Rev. Lett.* **115**, 017701 (2015).
87. del Hougne, P., Lemoult, F., Fink, M. & Lerosey, G. Spatiotemporal Wave Front Shaping in a Microwave Cavity. *Phys. Rev. Lett.* **117**, 134302 (2016).
88. del Hougne, P., Fink, M. & Lerosey, G. Shaping Microwave Fields Using Nonlinear Unsolicited Feedback: Application to Enhance Energy Harvesting. *Phys. Rev. Applied* **8**, 061001 (2017).
89. del Hougne, P., Imani, M. F., Fink, M., Smith, D. R. & Lerosey, G. Precise Localization of Multiple Noncooperative Objects in a Disordered Cavity by Wave Front Shaping. *Phys. Rev. Lett.* **121**, 063901 (2018).
90. del Hougne, M., Gigan, S. & del Hougne, P. Deeply Subwavelength Localization with Reverberation-Coded Aperture. *Phys. Rev. Lett.* **127**, 043903 (2021).
91. del Hougne, P., Fink, M. & Lerosey, G. Optimally diverse communication channels in disordered environments with tuned randomness. *Nat. Electron.* **2**, 36–41 (2019).
92. del Hougne, P., Davy, M. & Kuhl, U. Optimal Multiplexing of Spatially Encoded Information across Custom-Tailored Configurations of a Metasurface-Tunable Chaotic Cavity. *Phys. Rev. Applied* **13**, 041004 (2020).
93. Frazier, B. W., Antonsen, T. M., Anlage, S. M. & Ott, E. Deep Wavefront Shaping: Intelligent Control of Complex Scattering Responses with a Programmable Metasurface. *arXiv:2103.13500 [physics]* (2021).
94. Bruck, R. *et al.* All-optical spatial light modulator for reconfigurable silicon photonic circuits. *Optica* **3**, 396 (2016).
95. Resisi, S., Viernik, Y., Popoff, S. M. & Bromberg, Y. Wavefront shaping in multimode fibers by transmission matrix engineering. *APL Photonics* **5**, 036103 (2020).
96. Ma, G., Fan, X., Sheng, P. & Fink, M. Shaping reverberating sound fields with an actively tunable metasurface. *Proc. Natl. Acad. Sci. USA* **115**, 6638–6643 (2018).
97. Tian, Z. *et al.* Programmable Acoustic Metasurfaces. *Adv. Funct. Mater.* **29**, 1808489 (2019).
98. Achilleos, V., Theocharis, G., Richoux, O. & Pagneux, V. Non-Hermitian acoustic metamaterials: Role of exceptional points in sound absorption. *Phys. Rev. B* **95**, 144303 (2017).
99. Sweeney, W. R., Hsu, C. W., Rotter, S. & Stone, A. D. Perfectly Absorbing Exceptional Points and Chiral Absorbers. *Phys. Rev. Lett.* **122**, 093901 (2019).
100. Suwunnarat, S. *et al.* Towards a Broad-Band Coherent Perfect Absorption in systems without Scale-Invariance. *arXiv:2103.03668 [physics]* (2021).
101. Wang, C., Sweeney, W. R., Stone, A. D. & Yang, L. Observation of coherent perfect absorption at an exceptional point. *Science* **373**, 1261–1265 (2021).
102. Ma, Q. *et al.* Smart metasurface with self-adaptively reprogrammable functions. *Light Sci. Appl.* **8**, 98 (2019).
103. Qian, C. *et al.* Deep-learning-enabled self-adaptive microwave cloak without human intervention. *Nat. Photonics* **14**, 383–390 (2020).

104. Tan, S. *et al.* All-optical computation system for solving differential equations based on optical intensity differentiator. *Opt. Express* **21**, 7008 (2013).

# Supplementary Materials for

# Meta-Programmable Analog Differentiator

Jérôme Sol[1], David R. Smith[2], and Philipp del Hougne[3*]

[1] INSA Rennes, CNRS, IETR - UMR 6164, F-35000, Rennes, France

[2] Center for Metamaterials and Integrated Plasmonics, Department of Electrical and Computer Engineering, Duke University, Durham, NC, 27708 USA

[3] Univ Rennes, CNRS, IETR - UMR 6164, F-35000, Rennes, France

* Correspondance to philipp.del-hougne@univ-rennes1.fr.

## Supplementary Note 1. Transfer Function of an Ideal Differentiator

In this supplementary note, we prove for completeness that the transfer function of an ideal differentiator is $H(\omega) = i(\omega - \omega_0)$. We also discuss the impact of an additional background phase drift with frequency on the filter's functionality. Finally, we provide the transfer function of an ideal second-order differentiator.

Let $e(t)$ be the function whose derivative is to be determined. This function $e(t)$ modulates a carrier signal of angular frequency $\omega_0$, yielding the input signal

$$E_{in}(t) = e(t)e^{i\omega_0 t}.$$

The desired output signal is thus

$$E_{out}(t) = \frac{de(t)}{dt}e^{i\omega_0 t}.$$

The Fourier transform of $E_{out}(t)$ is

$$\tilde{E}_{out}(\omega) = \int_{-\infty}^{\infty} \frac{de(t)}{dt}e^{-i(\omega-\omega_0)t}dt.$$

Integration by parts yields

$$\tilde{E}_{out}(\omega) = e^{-i(\omega-\omega_0)t}e(t)\Big|_{-\infty}^{\infty} + i(\omega - \omega_0)\int_{-\infty}^{\infty} e(t)\,e^{-i(\omega-\omega_0)t}dt = i(\omega - \omega_0)\tilde{E}_{in}(\omega),$$

where we assume that $\lim_{t\to\pm\infty} e(t) = 0$ for any realistic signal $e(t)$.

The transfer function associated with the differentiation operation is thus

$$H(\omega) = \frac{\tilde{E}_{out}(\omega)}{\tilde{E}_{in}(\omega)} = i(\omega - \omega_0).$$

An analog differentiator is in general satisfactory if its output is proportional (not necessarily equal) to $E_{out}(t)$, i.e., $H(\omega) \propto i(\omega - \omega_0)$ is needed. In practice, we observe that instead we achieve $H(\omega) \propto i(\omega - \omega_0)e^{-i\tau\omega}$ due to a slow background phase drift with frequency. The basic time-shifting property of the Fourier transform, namely that the Fourier transform of $X(t - t_0)$ is $e^{-i\omega t_0}\tilde{X}(\omega)$, implies that this background phase drift $e^{-i\tau\omega}$ just shifts the desired output signal in time from $E_{out}(t)$ to $E_{out}(t - \tau)$, and hence poses no problem for the analog differentiator.

The transfer function of higher-order derivatives is simply the product of several first-order differentiator transfer functions, e.g. for the second derivative, $\frac{d^2 y}{dt^2} = \frac{d}{dt}\frac{dy}{dt}$, the corresponding transfer function is $|H(\omega)|^2 = -(\omega - \omega_0)^2$. The phase of an ideal second-order differentiator is thus perfectly flat.

## Supplementary Note 2. Vulnerability of Analog Differentiators

In this supplementary note, we illustrate the vulnerability of analog differentiators in order to highlight why it is important to *perfectly* implement the transfer function derived in Supplementary Note 1 in the experiment. Specifically, we show that an experimental setup does not faithfully act as differentiator if it

(i) relies on a well-matched port (e.g., at $-30$ dB) as opposed to a perfectly matched port,

(ii) gets the central operating frequency slightly wrong, or

(iii) presents an asymmetry around the central frequency.

To substantiate these points, without loss of generality, we take the analytical transfer function of a differentiator based on a simple Mach-Zehnder interferometer (MZI, akin to Refs.[1,2]) as starting point in order to work with a physically justified analytical expression. Assuming that transmission through the upper MZI arm, relative to the lower MZI arm, introduces an additional time delay $\tau$ and differs in amplitude by a factor $A$, then the transfer function reads $H(\omega) = 1 + Ae^{i\omega\tau}$. Under idealistic conditions, $A = 1$ and one would choose $\tau = \frac{(2m+1)\pi}{\omega_0}$, where $m$ is an arbitrary integer, such that $H(\omega_0) = 0$ and the desired transfer function from Supplementary Note 1 would be obtained in the vicinity of $\omega_0$ (see Refs.[1,2]).

In column **a** of Supplementary Figure 1, we depict the ideal case in which the transfer function perfectly matches the required one (see Supplementary Note 1) in terms of magnitude and phase, and the output signal upon injection of a Gaussian envelope is perfectly symmetric and has exactly zero amplitude at its center.

In column **b**, we introduce a minimal error by setting $A = 0.9975$ instead of unity. The magnitude of the transfer function thereby deteriorates to $-33.7$ dB at $\omega_0$ which is not even visible on the linear magnitude plot in the first row. However, it is clearly visible that the phase jump is not abrupt anymore. The output signal already displays substantial distortions: it is not symmetric anymore in the time domain and there is a non-zero component at $\omega_0$ in its spectrum. This highlights that a well-matched port ($-33.7$ dB) is not good enough to construct a faithful differentiator. In column **c**, we use $A = 0.99$ which makes the observations from **b** even more prominent.

In column **d**, we investigate the impact of deviating from the ideal case through a slight increase of the ideal value of $\tau$ by a factor of 1.01. This minimal change significantly shifts the frequency at which the zero lies on the real frequency axis to the left, such that the output signal has no resemblance with the analytically expected derivative of the Gaussian input envelope.

In column **e**, we investigate the role of asymmetry of the transfer function around $\omega_0$. We alter the ideal transfer function by multiplying it with a factor of 0.75 for $\omega < \omega_0$. The output signal is again significantly distorted: it clearly does not have zero amplitude at its center in the time domain anymore, and its magnitude in the spectral representation is clearly asymmetric now.

These illustrative examples highlight the importance of *perfectly* placing a zero of the scattering matrix exactly on the real frequency axis and exactly at the desired frequency of operation.

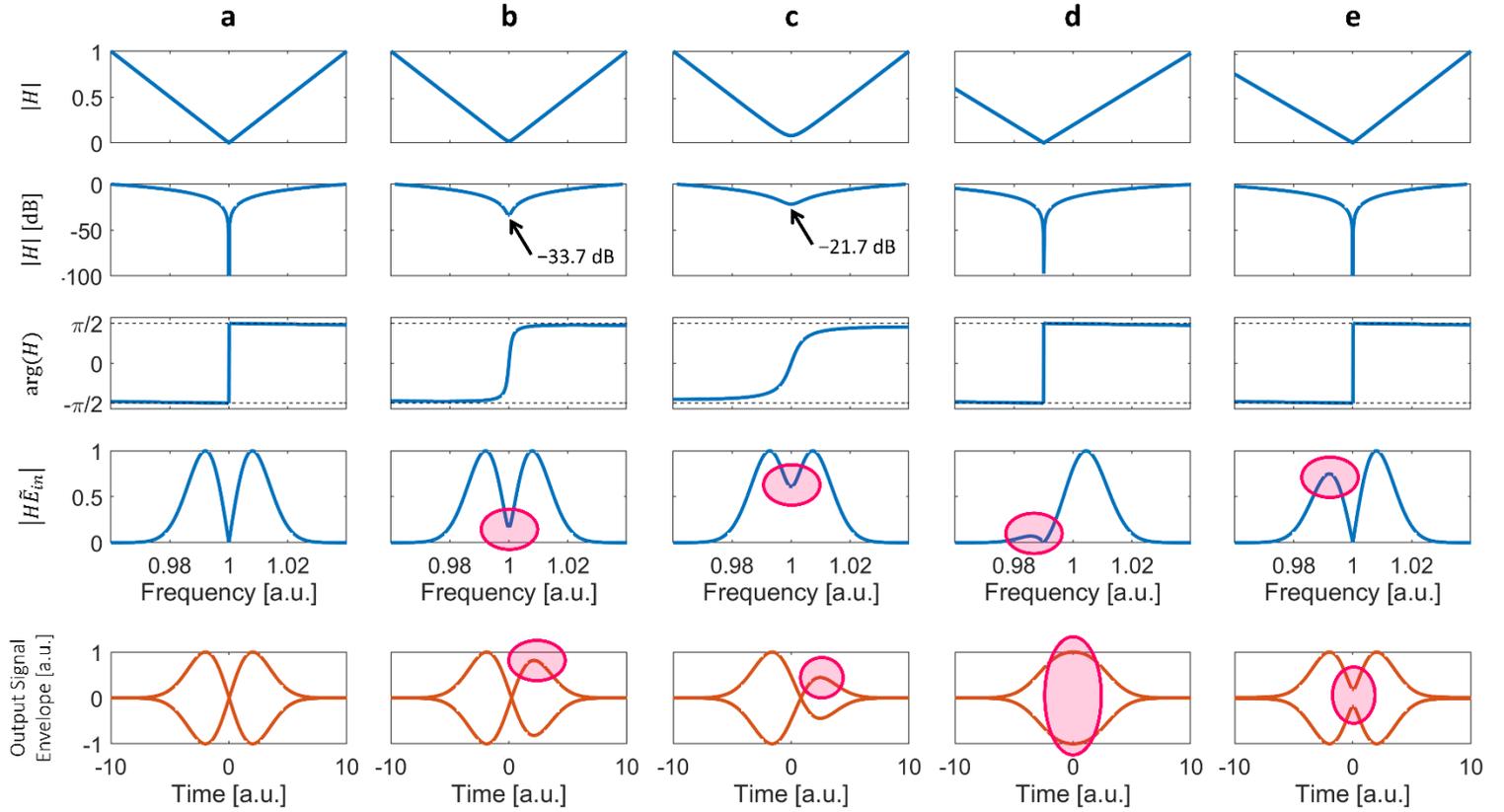

**Supplementary Figure 1. Vulnerability of analog differentiators.** For different transfer functions (**a-e**), the top row shows the magnitude of the transfer function on a linear scale, the second row shows the magnitude of the transfer function on a logarithmic scale, the third row shows the phase of the transfer function, the forth row shows the magnitude of the output signal's spectrum (for a Gaussian envelope as input signal), and the bottom row shows the envelope of the output signal (again for a Gaussian envelope as input signal). The different transfer functions correspond to: **a,** ideal; **b,** like ideal but with $A = 0.9975$; **c,** like ideal but with $A = 0.99$; **d,** like ideal but $\tau$ increased by a factor of 1.01; **e,** like ideal but amplitude of the transfer function for $\omega < \omega_0$ multiplied by a factor of 0.75.

## Supplementary Note 3. Details on the Experimental Setup

In this supplementary note, we provide further details on the experimental setup, including characterizations of the chaotic cavity in terms of its reverberation time and of the programmable metasurface in terms of its operating bandwidth.

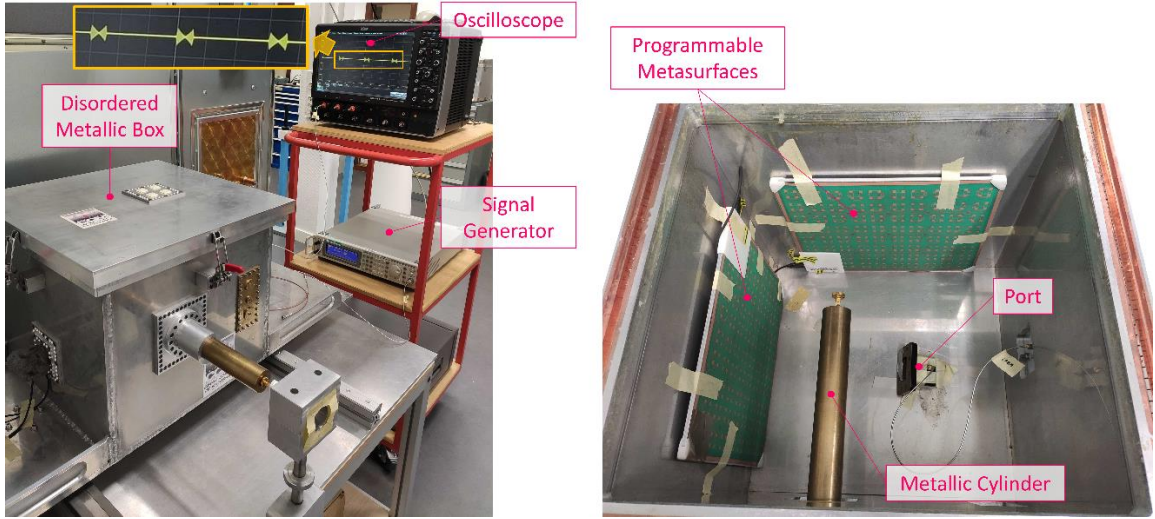

**Supplementary Figure 2. Photographic image of the experimental setup.** (right) Overview of experimental setup involving the disordered metallic box, the signal generator (Aeroflex IFR 3416, 250kHz-6GHz) and the oscilloscope (SDA 816Zi-B 16GHz Serial Data Analyzer, 80GS/s). In the pictured instance, the signal generator injects a waveform with quadratic envelope (signal power: 2dBm; carrier frequency: 5.1GHz) into the system, and the corresponding output signal is seen on the oscilloscope's display: a linear envelope. (left) Inside view of the metallic box with the top cover removed. Two programmable metasurfaces are placed on perpendicular walls, and a waveguide-to-coax adapter (RA13PBZ012-B-SMA-F) acts as port. A metallic cylinder breaks the symmetries of the box and ensures that the rays therein display chaotic behavior.

A photographic image of the setup corresponding to Figure 1 from the main text is shown in Supplementary Figure 2. The metallic box has dimensions of $0.385\text{m} \times 0.422\text{m} \times 0.405\text{m}$ and hence a volume of $0.0658\text{m}^3$. Chaotic behavior of the waves inside the metallic box is ensured via various mechanisms: on the one hand, the presence of metallic elements (notably the cylinder but also the port itself as well as irregularities on the wall) breaks the symmetries of the box; on the other hand, the use of metasurface configurations in which the meta-atoms are not all in the same state acts as an electronic equivalent of further geometric perturbations[3]. We determine the enclosure's quality factor as $Q = 410$ via the average decay rate of inverse Fourier transformed spectra measured for a series of random metasurface configurations. Based on Weyl's law, this means that at a given frequency within the considered interval $N \sim \frac{8\pi V}{c^3 Q} f_0^3 = \frac{8\pi}{Q} \frac{V}{\lambda_0^3} = 21$ modes overlap.

The programmable metasurface is an ultrathin array of meta-atoms whose electromagnetic scattering properties can be reconfigured electronically. The concept can be traced back to pioneering works in the early 2000s[4,5] and received renewed attention in 2014[6,7]. Since then, many designs for programmable metasurfaces have been proposed, and such devices are also known as "tunable impedance surface", "spatial microwave modulator" or "reconfigurable intelligent surface". Our present work could be implemented with any programmable metasurface design because it solely relies on the ability to somehow tune the complex scattering system's properties but not on the specific meta-atom properties. An ideal programmable metasurface for our purpose (i) interacts with as many rays as possible, meaning that each meta-atom has the largest possible scattering cross-section and the metasurface consists of as many meta-atoms as possible, (ii) the meta-atom programmability is as fine-grained as possible (but at least 1-bit), and (iii) insertion of the metasurface into the chaotic enclosure does not significantly alter the amount of absorption.

The prototype used in our experiments (purchased from Greenerwave) is based on the design introduced in Ref.[7] and has previously been used in Refs.[3,8–12]. In our setup seen in Figure 1 of the main text and Supplementary Figure 2, the programmable metasurfaces cover 16.2% of the cavity's wall surface. In the setup seen in Figure 4a of the main text, the programmable metasurface covers 8.1% and 7.2% of the wall surface area in the two cavities, respectively. Each meta-atom (see Supplementary Figure 3a) has two digitalized states, "0" and "1", and can be toggled between these two states by controlling the bias voltage of an integrated PIN diode. Specifically, each meta-atom consists of two resonators that hybridize, as detailed in Ref.[7], and via the bias voltage of a PIN diode the resonance frequency of one of the two resonators can be altered. Thereby, the phase change of the reflected wave can be tuned by roughly $\pi$. The meta-atoms in Ref.[7] acted on a single field polarization, whereas the meta-atoms of the prototype that we utilize can be thought of as the fusion of two such meta-atoms, one rotated by 90°, each acting on one polarization of the electromagnetic field.

We begin by characterizing the utilized programmable metasurface in the conventional way, namely to consider the phase shift of the reflected wave between the two possible states for normally incident waves. The metasurface consists of 76 of the above-described meta-atoms. To that end, we utilize the horn-antenna setup shown in Supplementary Figure 3a and synchronize the states of all meta-atoms for this measurement. We measure the return loss of the horn antenna when the meta-atoms are all simultaneously in their two possible states: "0" or "1". The magnitude of the return loss in the two cases, as well as the phase difference of the return loss between the two cases, are plotted in Supplementary Figure 3b,c. The return loss magnitudes are of course modulated by the horn antenna's transfer function and thus not a quantification of the energy that is absorbed by the metasurface, but it is apparent that there is no significant difference in terms of reflected field magnitude between the two states. The phase difference reaches the ideal value of $\pi$ in the vicinity of 5.15 GHz. In other words, at this frequency, every meta-atom can be configured to mimic Dirichlet or Neuman boundary conditions. Similar results were measured for the other field polarization.

The above "conventional" characterization is a useful first indication of the metasurface's properties but of course it contains no information about the dependence of these properties on the angle of incidence or on the coupling between different meta-atoms. Especially in a rich scattering setting such as the one we consider, in which certainly waves from all possible angles are incident on the metasurface, it is more meaningful to characterize the metasurface *in situ*[13,14]. To that end, we place the two programmable metasurfaces at their intended locations, as seen in Supplementary Figure 3d, and measure the scattering parameter of interest (here $S_{11}(f)$) for a series of 500 random metasurface configurations. By evaluating the standard deviation across these measurements, we obtain a useful metric to assess the extent to which the programmable metasurface impacts the considered scattering parameter in the considered setup. Of course, this metric is still somewhat influenced by the specific system, so we also plot a smoothed version of the curve, obtained with a sliding average filter. Overall, it can be seen that the programmable metasurface most efficiently modulates the field within a 400 MHz bandwidth centered roughly on 5.15 GHz.

For the experiment underlying Figure 4 in the main text, we use an additional second metallic box of dimensions $0.5\text{m} \times 0.5\text{m} \times 0.3\text{m}$ and hence of volume of $0.075\text{m}^3$.

Chaotic behavior of the waves inside the metallic box is ensured via various mechanisms: on the one hand, the presence of metallic elements (notably two metallic hemispheres but also the port itself as well as irregularities on the wall) breaks the symmetries of the box; on the other hand, the use of metasurface configurations in which the meta-atoms are not all in their perfect electric conductor (PEC) equivalent state acts as an electronic equivalent of further geometric perturbations[3]. We determine the enclosure's quality factor as $Q = 446$ via the average decay rate of inverse Fourier transformed spectra measured for a series of random metasurface configurations. Based on Weyl's law, this means that $N \sim \frac{8\pi V}{c^3 Q} f_0^3 = \frac{8\pi}{Q} \frac{V}{\lambda_0^3} = 22$ modes overlap at a given frequency within the considered interval.

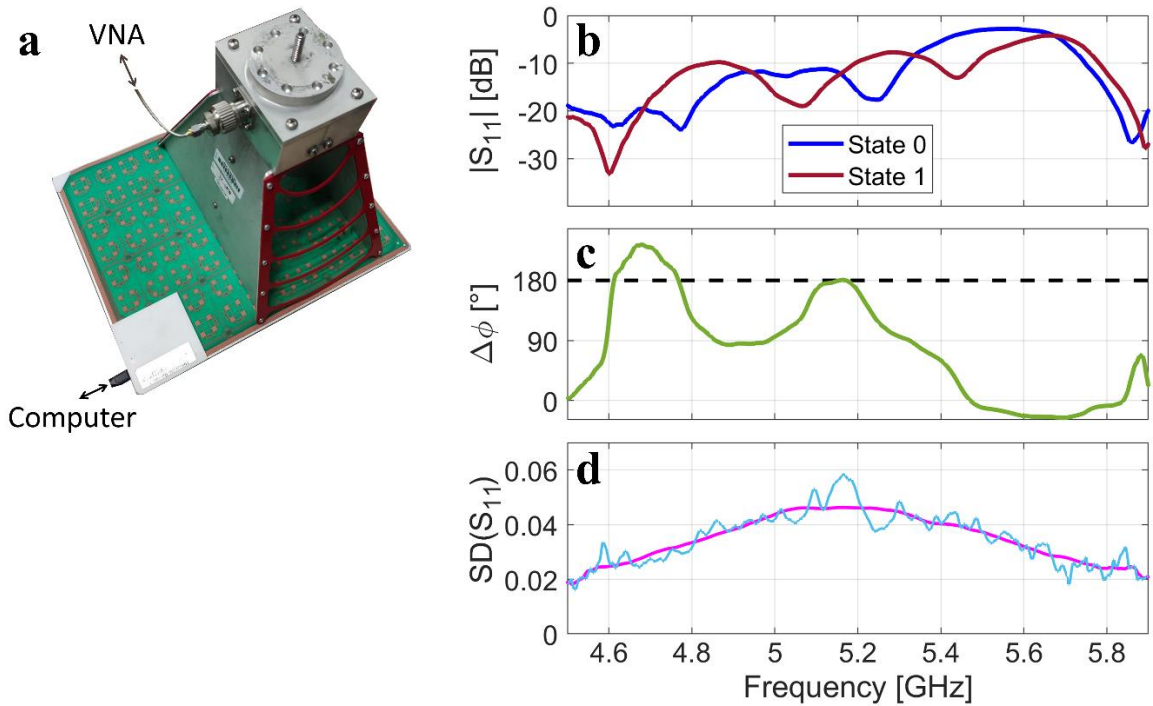

**Supplementary Figure 3. Characterization of the programmable metasurface. a,** Setup to characterize the response under normal incidence. **b,** Magnitude of the return loss measured with the setup from **a** when all meta-atoms are simultaneously either in state "0" or state "1". **c,** Phase difference between the return losses measured with the setup in **a** in the two possible states. **d,** *In situ* characterization of the metasurface via the standard deviation of the reflection spectrum measured inside the disordered metallic box for 500 random metasurface configurations. A smoothed version of the curve is also shown.

**Supplementary Note 4**. **Details on the Experimental Procedure**

In this supplementary note, we discuss procedural details of our experiments, including the identification of suitable metasurface configurations, the direct observation of temporal differentiation of various waveforms, as well as details regarding the modified setups underlying Figures 3 and 4 of the main text.

In order to separate injected and reflected waveform, port 2 of a circulator (PE83CR006) is connected to the coax-to-waveguide adapter inside the metallic disordered box, as seen in Supplementary Figure 4.

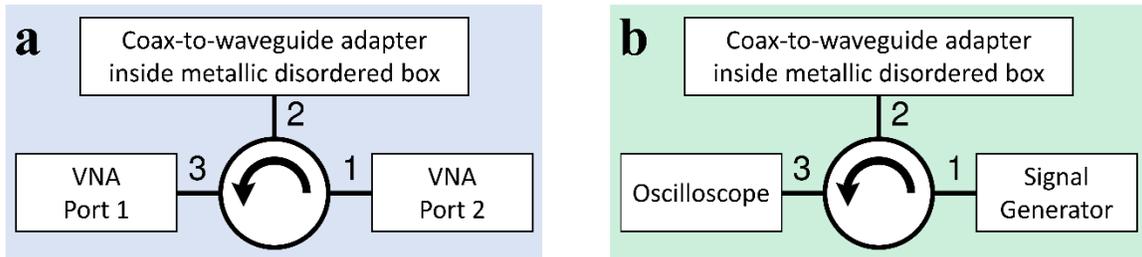

**Supplementary Figure 4.** Use of a circulator to separate injected and reflected signals. **a,** Port connections involving a VNA used to optimize the metasurface configuration. **b,** Port connections involving a signal generator and oscilloscope used to perform temporal differentiation.

Identifying a metasurface configuration that yields a zero at a desired frequency is not a trivial task because no analytical or learned forward model is available that would allow us to predict the scattering response which corresponds to a given metasurface configuration. Given that the amount of degrees of freedom in the present work is one order of magnitude larger than in Ref.[15], measuring the scattering response for all possible metasurface configurations is intractable. Instead, we use an iterative experimental trial-and-error optimization method. To that end, we connect a vector network analyzer (Agilent Technologies PNA-L Network Analyzer N5230C, 300kHz-20GHz) to ports 1 and 3 of the circulator, as shown in Supplementary Figure 4a. We operate with an emitted power of 0 dBm and an IF bandwidth of 10 kHz. The transmission from port 1 to port 3 of the circulator is the reflection off the port inside the disordered metallic box. First, we measure the scattering response for 100 random metasurface configurations. We pick the one closest to our objective as starting point. Then, we flip the state of one meta-atom at a time; we keep the change if the resulting scattering response is closer to our objective. We observe that typically after a maximum of roughly 700 iterations no further improvement is observed. Like most inverse design methods, ours does not guarantee the identification of the globally optimal metasurface configuration. However, we observe that different optimization runs yield outcomes of comparable quality. In the future, we are confident that learned forward models based on ANNs can be implemented, such that the need for this iterative experimental optimization can be circumvented. Ref.[16] already employs a learned forward model to predict the scattering response of a programmable metasurface, albeit in quasi free space rather than inside a complex scattering enclosure. Such ANN-based approaches will greatly benefit from next-generation meta-atoms with fine-grained programmability ($>$ 1-bit) because they become compatible with continuous gradient-descent optimization protocols. Note that in certain settings without environmental perturbations during runtime, the identification of suitable metasurface configurations can be completed offline during a calibration phase and presents no burden during runtime.

To directly observe the wave-based analog computation of temporal derivatives, we generate various waveforms with a signal generator and observe the reflected signals on an oscilloscope, with connections as shown in Supplementary Figure 4b. Specifically, for the main experiments we use a signal generator (Aeroflex IFR 3416, 250kHz-6GHz) to generate a signal $e(t)e^{i\omega_0 t}$. An arbitrary signal envelope $e(t)$ is defined and sampled at 33 MHz, and an arbitrary value of $\omega_0$ can be chosen within the considered 5-GHz-band. We emit the signals at 2 dBm. The emitted signal envelopes (GAUSSIAN, QUADPOLY, SKYLINE) are displayed in Figures 2c,f,i in the main text. The signal envelopes are repeated in intervals of 3 µs. The oscilloscope (SDA 816Zi-B 16GHz Serial Data Analyzer 80GS/s) measures a 10 µs interval of the reflected signal with a sampling rate of 40 GS/s.

In order to inject the sum of two signals $e_A(t)e^{i\omega_A t} + e_B(t)e^{i\omega_B t}$ for the experiments underlying Figure 3 of the main text, we use a second signal generator (Agilent MXG Analog N5183A, 100kHz-20GHz). Signal generators A and B generate $e_A(t)e^{i\omega_A t}$ and $e_B(t)e^{i\omega_B t}$, respectively, and the two generated signals are then summed using a simple "T" connector before being injected into the port. The second signal generator generates a square wave with period 2.5 µs at -13 dBm and an arbitrary value can be chosen for $\omega_0$.

The reason for emitting at such low power is that the peaks of the square wave's derivatives are very large and we intend to measure them with the same dynamic range as the derivatives of the other signal envelopes (GAUSSIAN, QUADPOLY, SKYLINE). The two signals from generators A and B are not synchronized in any manner and have different repetition intervals.

To suppress measurement noise, all measured data is digitally filtered to impose a pass-band of $f_0 \pm 0.02 \text{ GHz}$, where $f_0$ is the carrier frequency. The reflected signal's amplitude is intrinsically low for a differentiator due to the zero at the carrier frequency in its transfer function. Hence, the output signal is particularly weak in the case of a second-order differentiator as considered in Figure 4 of the main text. Therefore, the output signal envelopes displayed in Figure 4 of the main text are averaged over 20 acquisitions in order to limit the corruption through measurement noise.

## Supplementary Note 5. Experimentally Measured Notch Depths

In this supplementary note, we provide further details on the experimentally measured notch depths for the data displayed in Figure 2a of the main text. To that end, we reproduce Figure 2a,b from the main text in Supplementary Figure 5 and add a plot of the transfer function magnitude on a logarithmic scale. Moreover, we summarize the notch depth for each considered central frequency in a table below.

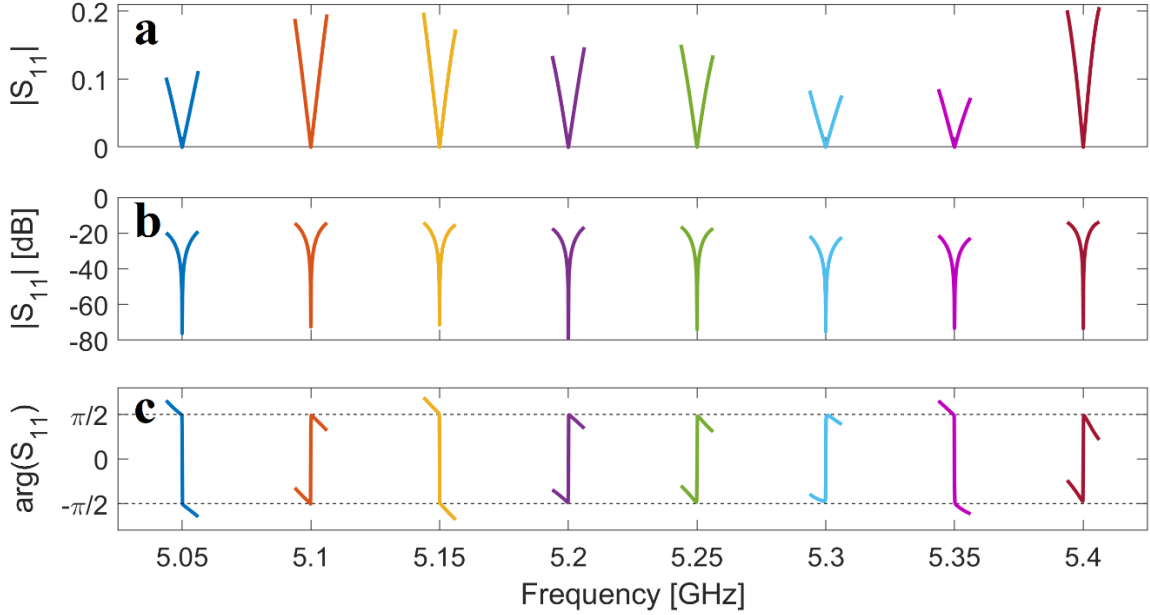

**Supplementary Figure 5. Experimentally measured notch depths.** Measured transfer functions in terms of magnitude on a linear scale (**a**), magnitude on a logarithmic scale (**b**), and phase (**c**). Subfigures **a** and **c** are reproduced from Figure 2a,b in the main text.

| Central Frequency [GHz] | Notch Depth [dB] |
|---|---|
| 5.05 | -76.3 |
| 5.10 | -72.5 |
| 5.15 | -71.4 |
| 5.20 | -82.0 |
| 5.25 | -74.1 |
| 5.30 | -75.3 |
| 5.35 | -73.3 |
| 5.40 | -73.5 |

## Supplementary Note 6. Error Performance Evaluation: Bandwidth and Dip Symmetry

In this supplementary note, we analyze the computation error of our meta-programmable differentiator based on a measured system transfer function. We explore the error dependence on the input signal's bandwidth and discuss the role of the symmetry of the reflection dip.

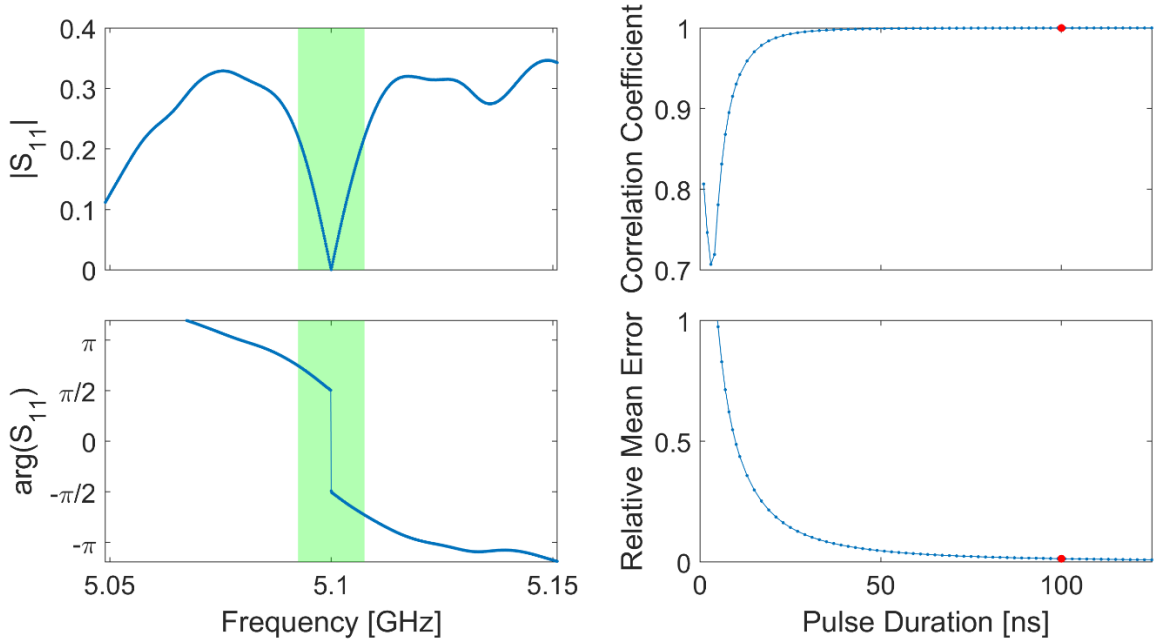

**Supplementary Figure 6. Evaluation of computation error as a function of input signal bandwidth. a,b,** Magnitude (**a**) and phase (**b**) of the measured transfer function with a zero at 5.1 GHz that is used for the error evaluation. The shaded area indicates the bandwidth of 15 MHz within which this transfer function is a good approximation to that of an ideal differentiator. **c,d,** Computational error as a function of input signal duration in terms of two metrics: correlation coefficient (**c**) and relative mean error (**d**). To compute these metrics, the output signal for a given input pulse duration is calculated, normalized and compared to the analytically expected output signal. The red dot indicates the duration of the Gaussian pulse used in the main text (Figure 2c).

The ideal differentiator's transfer function is $H(\omega) = i(\omega - \omega_0)$ – see Supplementary Note 1; its magnitude is hence linear and symmetric with respect to $\omega_0$. Any physical implementation of this transfer function is only an approximation within some bandwidth around $\omega_0$ and hence intrinsically bandwidth-limited. Most wave-based differentiators that have been proposed so far (see references in the main text's introduction) are based on regularly shaped devices and their scattering matrix often only has a single zero within the considered frequency range. In contrast, our overmoded random scattering system has typically many other poles and zeros (with non-zero imaginary component) in the vicinity of the zero that we tune to a real frequency. These are, of course, in principle not distributed symmetrically around the targeted frequency and can hence introduce strong asymmetries outside the direct vicinity of the real-valued zero. Therefore, our optimization protocol not only minimizes $|H(\omega_0)|$ but also pays attention to the symmetry of $|H(\omega)|$ in the vicinity of $\omega_0$. Indeed, different locally optimal metasurface configurations that minimize $|H(\omega_0)|$ have been found to sometimes vary drastically in terms of symmetry and thus useful bandwidth. The stronger the symmetry is, the larger is the maximum input signal bandwidth for which an acceptably low computation error can be guaranteed.

Considering the case of an ideal Gaussian pulse as input (similar to Figure 2c in the main text), we calculate the (normalized) output signal envelope and compare it to the analytically expected output signal envelope for an ideal differentiator using two metrics: the relative mean difference between the two, as well as the correlation coefficient of the two. This calculation is performed for various bandwidths of the input pulse. The results in Supplementary Figure 6 show that up to an input signal bandwidth of roughly 15 MHz our system's performance is very close to that of an ideal differentiator.

In the table below, we compare the fractional bandwidth achieved by various wave-based temporal differentiators reported in the literature. Refs.[17–20] are static (not programmable) wave-based photonic differentiators, while Ref.[2] is a reconfigurable integrated interferometric photonic differentiator. None of these devices experiences symmetry issues regarding the transfer function magnitude (as discussed above) but implementation-specific effects eventually cause deviations from the ideal linear shape; for instance, in Ref.[17] due to a slight non-linear dispersion slope of the core and cladding modes. In terms of fractional bandwidth, a quantity that is independent of the central operating frequency and hence most meaningful from a general wave engineering perspective, our device achieves a performance of the same order of magnitude as Ref.[20] and only Ref.[17] performs significantly better. Refs.[17,20] are both static and the only report on a somewhat tunable differentiator in Ref.[2] has a fractional bandwidth which is an order of magnitude lower than the one reported in this present work. Interestingly, Refs.[2,19] observed that the computational error deteriorates if the input bandwidth gets too large *or too small* – unlike our work in which only too large signal bandwidths appear to deteriorate the computational precision. Finally, we note that our meta-programmable analog differentiator's operating bandwidth of 15 MHz is of the same order of magnitude as the channel width of typical WLAN wireless communication channels operating at the same central frequency – paving the path to direct applications of our technique in this area.

| Reference | Operation Principle | Central Frequency $f_0$ | Frequency Bandwidth $\Delta f$ | Fractional Bandwidth $\Delta f / f_0$ |
|---|---|---|---|---|
| Ref.[17] | Long-period fiber grating. | 195 THz | 2.3 THz | $1.5 \times 10^{-2}$ |
| Ref.[18] | Microring resonator. | 193 THz | 42 GHz | $2 \times 10^{-4}$ |
| Ref.[19] | Fiber Bragg grating. | 193.5 THz | 25 GHz | $1 \times 10^{-4}$ |
| Ref.[20] | Directional coupler. | 193.5 THz | 1.25 THz | $6.5 \times 10^{-3}$ |
| Ref.[2] | Tunable interferometer. | 192 THz | 55 GHz | $3 \times 10^{-4}$ |
| This work. | Tuned overmoded chaotic cavity. | 5.1 GHz | 15 MHz | $3 \times 10^{-3}$ |

## Supplementary Note 7. Operation in Transmission Mode

In the main text, we presented results obtained in reflection mode. In this supplementary note, we present complementary results obtained in transmission mode for a first-order temporal differentiator. In reflection mode, a single port is used but a circulator is necessary to separate incident and reflected signal (see Supplementary Figure 4); in transmission mode, two ports are required but there is no need for a circulator.

The main inconvenience of operating in transmission mode is the significantly weaker magnitude of the transfer function in comparison to operating in reflection mode; in our experiments, the transfer function magnitude, averaged over random metasurface configurations and all frequency points, is 17.74 dB lower if operating in transmission:

$\langle |S_{11}| \rangle = -15.22$ dB in contrast to $\langle |S_{21}| \rangle = -27.96$ dB. This difference in transfer function magnitude results in a drastically higher sensitivity to noise during the optimization of the metasurface configuration, as well as less energy-efficient analog wave processing, if operation in transmission mode is chosen.

Nonetheless, V-shaped transmission zeros can be imposed on demand analogously to the V-shaped reflection zeros discussed in the main text, and the resulting transfer function is that of a first-order differentiator. Refs.[13,21] previously minimized the transmission between two ports in comparable metasurface-tunable complex scattering settings to create "cold spots", but these works did not implement true transmission zeros. An example of a transmission zero that we implemented in our transmission-mode setup is shown in Supplementary Figure 7 below. In comparison to the reflection-mode setup detailed in Figure 1 in the main text, we removed the circulator and instead added a second port inside the chaotic cavity. The second port is identical to the first port and placed at least half a wavelength away from the first port, in perpendicular orientation to the first port. The transfer function's magnitude has the desired linear behavior around its minimum at $-79.5$ dB and the phase displays the required $\pi$ phase jump across the central frequency. Again, a global phase drift is seen, which does not impact the differentiator functionality (see Supplementary Note 1). Upon injection of a Gaussian pulse (same parameters as in the main text), the corresponding output measurement yields the curve displayed in Supplementary Figure 7d. The principal features of the analytically expected output signal envelopes are observed (two symmetric pulses and a zero between them), but the quality of the differentiation operation is clearly worse than that reported in the main text in reflection mode. The very low signal strength in comparison to the noise floor is apparent; neater results could be obtained by averaging over multiple acquisitions.

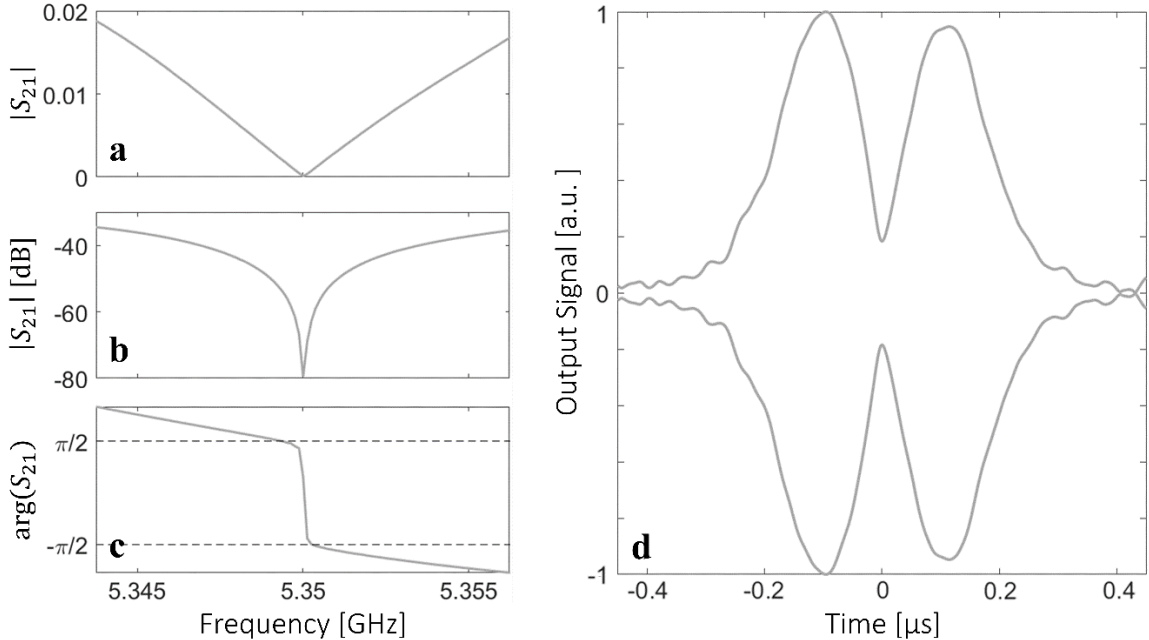

**Supplementary Figure 7. Example of an analog differentiator operating in transmission mode at 5.35 GHz. a,b,c,** The left column displays the transfer function, here $S_{21}(f)$, in terms of its absolute value on a linear (**a**) and logarithmic (**b**) scale, as well as the associated phase (**c**). **d,** Normalized envelope of the measured output signal (upon injection of a Gaussian pulse).

## Supplementary Note 8. Further Examples of Parallel Computing

In the main text, we provided in Figure 3 several examples of computing two derivatives simultaneously by imposing simultaneously two zeros on the transfer function at distinct frequencies $\omega_A$ and $\omega_B$. We reported direct evidence of parallel computing by injecting the sum of two independent envelopes modulated onto carriers $\omega_A$ and $\omega_B$, respectively. In this supplementary note, we provide further examples of measured optimized transfer functions for parallel differentiation, for up to four independent data streams. Given that only two signal generators were at our disposal, we could not directly test these further examples of parallel computing which would require the simultaneous generation of three or four independent waveforms.

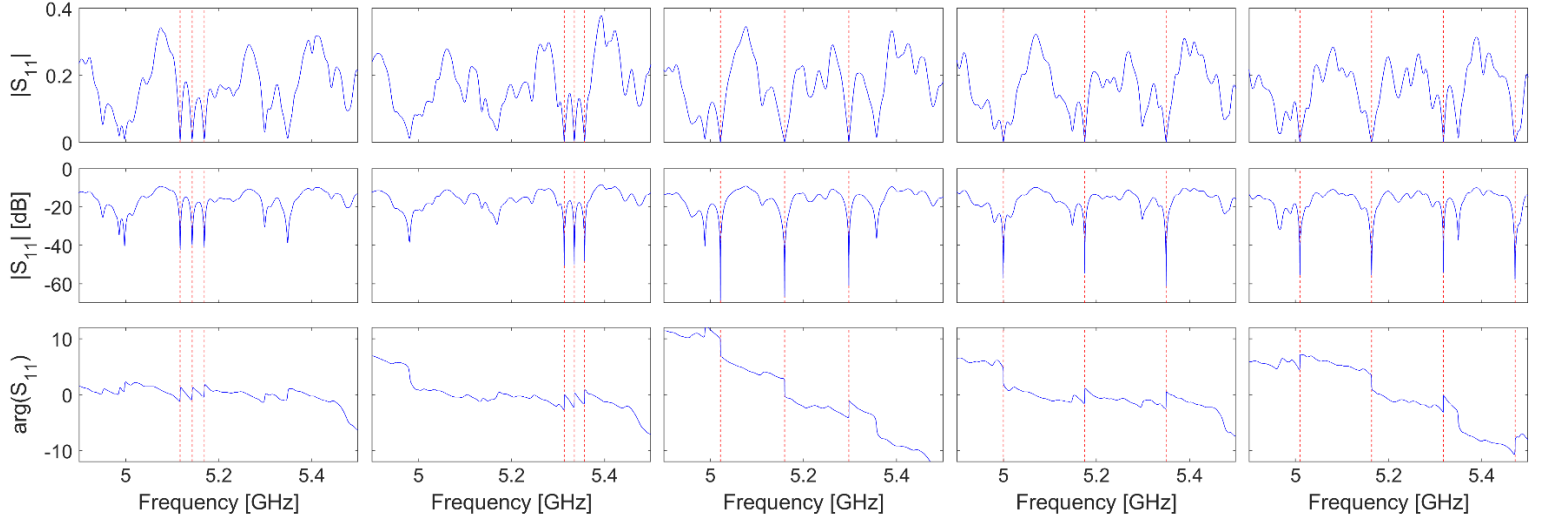

**Supplementary Figure 8. Various examples of optimized transfer functions with multiple zeros for parallelized wave-based differentiation.** For each example, the transfer function's amplitude is shown on a linear scale (top row) and on a logarithmic scale (middle row), as well as the corresponding phase (bottom row). Four examples with three scattering zeros and one example with four scattering zeros are shown.

Five examples of experimentally optimized and measured transfer functions with three or four real-valued scattering matrix zeros in the considered 5-GHz-band are displayed in Supplementary Figure 8. The regular spacing of these zeros was chosen on purpose. It is apparent that closely spaced zeros are more difficult to impose (the first two examples in Supplementary Figure 8). However, in all shown cases the reflection dips are sufficiently deep to serve for the desired differentiation functionality. An example with four zeros is also shown in Supplementary Figure 8 (last example). In principle, such multi-objective optimizations are more demanding than imposing a single zero. To ease the optimization burden, depending on the requirements of the specific intended application it may be possible to relax constraints (for instance, to only fix the zeros' spacing but not their exact positions; to only fix the number of desired zeros; etc.). Alternatively, more degrees of freedom by using a larger programmable metasurface can improve the ability to satisfy all constraints.

Finally, we point out that parallel computing as reported in our work based on an overmoded chaotic cavity with many tunable degrees of freedom is not possible with the same flexibility if a tunable interferometer is used instead as physical device, as in Ref.[2].

Indeed, in the latter case, the spacing of carrier frequencies must be equal to the free spectral range and cannot be imposed at will – see, for instance, Figure 3b in Ref.[2].

**Supplementary Note 9**. **Generalizations**

In the main text, we presented meta-programmable analog differentiation of temporally encoded signals through "over the air" wave propagation in a programmable complex scattering enclosure in the microwave domain. In this supplementary note, we provide details on how our proposed concept can be generalized to (i) *spatially* encoded information, (ii) programmable overmoded random scattering system based on *guided* waves, and (iii) *acoustic or optical* scattering.

(i) Spatially Encoded Information

To obtain the hallmark V-shaped transfer function in our work, as seen for instance in Figure 2, we impose a zero for the chosen set of parameters at which we desire critical coupling (in particular, $\omega_0$); then, we trace the transfer function upon continuous detuning of the frequency, yielding the V-shape in the vicinity of the critical coupling condition. Thus, information that is temporally encoded into the input signal will be subject to the differentiation transfer function.

However, the same V-shaped spectrum is also obtained in the vicinity of the critical coupling condition by detuning any other parameter. Therefore, if the input signal is not incident through a mono-modal coaxial waveguide but carried by a wave propagating in free space that impinges at some angle $\theta$ on a perforated wall through which it couples to our programmable chaotic cavity, as sketched in Supplementary Figure 9, we can tune the system once again to critical coupling for a chosen set of parameters (in particular, $\theta_0$); then, we can trace the transfer function upon continuously detuning the angle of incidence, yielding once again the V-shape in the vicinity of the critical coupling condition. This time, however, the horizontal axis will be the angle of incidence as opposed to the frequency. Thus, information that is spatially encoded into the input signal (over a range of incidence angles) will be subject to the differentiation transfer function.

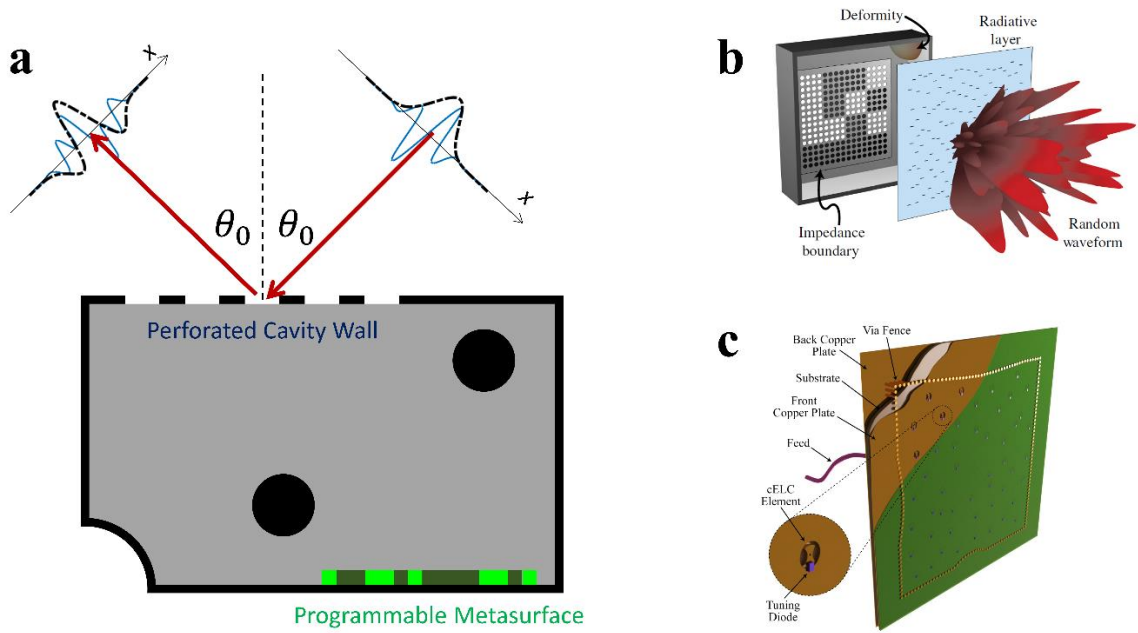

**Supplementary Figure 9. Metaprogrammable *spatial* analog differentiator. a,** Concept. **b,** Implementation in a 3D chaotic cavity from Ref.[22]. **c,** Implementation in a flat quasi-2D chaotic cavity from Ref.[23].

Therefore, our technique can straightforwardly be extended to meta-programmable spatial (as opposed to temporal) analog differentiation, and various technique to perforate a wall of 3D or quasi-2D chaotic cavities are well-known from setups that are routinely used for computational meta-imaging, e.g., in Refs.[22,23]. The optimization to tune a zero of the scattering matrix to the real frequency axis for the desired set of parameters (central frequency, angle of incidence) would be exactly the same as that used in the work we presented.

However, the differentiation of spatially as opposed to temporally encoded information appears more relevant to optical signal processing than to the microwave domain which is targeted in our work. In the microwave domain, information is temporally encoded into many signals of interest, for instance, in wireless communication or radar.

(ii) <u>Programmable Overmoded Random Scattering System Based on Guided Waves</u>

At the core of our technique lies the idea to purposefully perturb an overmoded random scattering system with many degrees of freedom, because such a system presents a high density of zeros that are statistically uniformly distributed along the real frequency axis. Therefore, our technique finds it easy to tune one or multiple of these zeros onto the real frequency axis at the desired position(s). Our experimental demonstration relies on a 3D metallic complex scattering enclosure because of its high practical relevance: The elegance of our experiments is that any metallic enclosure (toolbox, microwave oven, etc.) can be taken and only an ultrathin programmable metasurfaces needs to be attached to the walls. In the context of upcoming 6G wireless communication, many rooms which are also scattering enclosures at $< 6\ \mathrm{GHz}$ will naturally be equipped with "reconfigurable intelligent surfaces" such that the entire hardware needed to implement our concept "*over the air*" is already there.

Nonetheless, our concept can also be implemented in programmable overmoded random scattering systems based on *guided* waves, such as a network of coupled transmission lines with complex connectivity (also known as graph)[24,25]. The programmability could then originate, for instance, from a series of phase shifters integrated into the connections between nodes of the graphs. Indeed, recently CPA has been experimentally observed also on chaotic graphs in Ref.[26] (albeit not "on-demand" as in our work in order to impose CPA at any desired real frequency). In contrast to our "over-the-air" experimental implementation, a guided approach would require one to carry around a network of transmission lines unless there was a safe method to leverage, for instance, the wiring inside a building that exists for electricity supply.

(iii) <u>Toward Implementations in Acoustic and Optical Scattering</u>

The experiments reported in our manuscript were conducted in the microwave domain not merely because this is a convenient regime for proof-of-principle experiments but because it is of direct technological relevance in sight of many examples of microwave signals carrying temporally encoded information, such as in wireless communication or radar. Nonetheless, the fundamental principles underlying our technique are generic to any wave scattering system and can hence be extended to different parts of the electromagnetic spectrum as well as to acoustic scattering, provided a suitable mechanism to implement individually programmable scatterers can be identified. In this section, we provide a succinct survey of recent works from the literature that implemented techniques which may be leveraged to implement our proposal, for instance, in acoustic or optical scattering.

To begin with, we note that programmable meta-atoms very similar to the microwave prototype we used can be implemented at much higher electromagnetic frequencies in the THz and even infrared regimes using several well-established switchable diodes. For example, Schottky diodes in the THz regime (see Ref.[27] and the website of Teratech Components Ltd: http://www.teratechcomponents.com/) and thermal $VO_2$ diodes in the infrared regime (see Refs.[28,29]) can be used. Detailed design proposals for programmable meta-atoms based on these switchable diodes for operation around 350 GHz and 150 THz, respectively, can be found in the electronic supplementary material of Ref.[30].

In the optical regime, programmable scatterers have been demonstrated based on electro-optical modulation[31], phase-change materials[32–34], applied-voltage-sensitive graphene[35–37], all-optical modification of the spatial refractive index profile[38], acousto-optic modulation[39], spatial light modulators (SLM, if the scattering system is defined as the ensemble of complex medium and SLM)[40], as well as mechanical actuation[41,42] and computer-controlled mechanical perturbations[43].

In the acoustic regime, programmable scatterers have been demonstrated based on magnetic-field controlled elastomers[44], tunable membranes[45–48], geometrically-tunable resonators[49–53], as well as electrolysis-controlled microbubble arrays for ultrasound[54].

The existence of such a variety of experimental reports on tunable scattering in optics and acoustics suggests that our concept can be implemented with relative ease for optical or acoustic scattering, too.

## Supplementary References

1. Park, Y., Azaña, J. & Slavík, R. Ultrafast all-optical first- and higher-order differentiators based on interferometers. *Opt. Lett.* **32**, 710 (2007).
2. Liu, W. *et al.* A fully reconfigurable photonic integrated signal processor. *Nat. Photonics* **10**, 190–195 (2016).
3. Gros, J.-B., del Hougne, P. & Lerosey, G. Tuning a regular cavity to wave chaos with metasurface-reconfigurable walls. *Phys. Rev. A* **101**, 061801 (2020).
4. Sievenpiper, D. F., Schaffner, J. H., Song, H. J., Loo, R. Y. & Tangonan, G. Two-dimensional beam steering using an electrically tunable impedance surface. *IEEE Trans. Antennas Propag.* **51**, 2713–2722 (2003).
5. Holloway, C. L., Mohamed, M. A., Kuester, E. F. & Dienstfrey, A. Reflection and Transmission Properties of a Metafilm: With an Application to a Controllable Surface Composed of Resonant Particles. *IEEE Trans. Electromagn. Compat.* **47**, 853–865 (2005).
6. Cui, T. J., Qi, M. Q., Wan, X., Zhao, J. & Cheng, Q. Coding metamaterials, digital metamaterials and programmable metamaterials. *Light Sci. Appl.* **3**, e218–e218 (2014).
7. Kaina, N., Dupré, M., Fink, M. & Lerosey, G. Hybridized resonances to design tunable binary phase metasurface unit cells. *Opt. Express* **22**, 18881 (2014).
8. del Hougne, P., Davy, M. & Kuhl, U. Optimal Multiplexing of Spatially Encoded Information across Custom-Tailored Configurations of a Metasurface-Tunable Chaotic Cavity. *Phys. Rev. Applied* **13**, 041004 (2020).
9. del Hougne, P., Savin, D. V., Legrand, O. & Kuhl, U. Implementing nonuniversal features with a random matrix theory approach: Application to space-to-configuration multiplexing. *Phys. Rev. E* **102**, 010201 (2020).
10. del Hougne, P., Yeo, K. B., Besnier, P. & Davy, M. Coherent Wave Control in Complex Media with Arbitrary Wavefronts. *Phys. Rev. Lett.* **126**, 193903 (2021).
11. del Hougne, P., Yeo, K. B., Besnier, P. & Davy, M. On-Demand Coherent Perfect Absorption in Complex Scattering Systems: Time Delay Divergence and Enhanced Sensitivity to Perturbations. *Laser Photonics Rev.* **15**, 2000471 (2021).
12. del Hougne, P. *et al.* Diffuse field cross-correlation in a programmable-metasurface-stirred reverberation chamber. *Appl. Phys. Lett.* **118**, 104101 (2021).
13. Kaina, N., Dupré, M., Lerosey, G. & Fink, M. Shaping complex microwave fields in reverberating media with binary tunable metasurfaces. *Sci. Rep.* **4**, 6693 (2015).
14. Alexandropoulos, G. C., Shlezinger, N. & del Hougne, P. Reconfigurable Intelligent Surfaces for Rich Scattering Wireless Communications: Recent Experiments, Challenges, and Opportunities. *IEEE Commun. Mag.* **59**, 28–34 (2021).
15. Imani, M. F., Smith, D. R. & del Hougne, P. Perfect Absorption in a Disordered Medium with Programmable Meta-Atom Inclusions. *Adv. Funct. Mater.* **30**, 2005310 (2020).
16. Li, H.-Y. *et al.* Intelligent Electromagnetic Sensing with Learnable Data Acquisition and Processing. *Patterns* **1**, 100006 (2020).
17. Slavík, R., Park, Y., Kulishov, M., Morandotti, R. & Azaña, J. Ultrafast all-optical

differentiators. *Opt. Express* **14**, 10699 (2006).
18. Liu, F. *et al.* Compact optical temporal differentiator based on silicon microring resonator. *Opt. Express* **16**, 15880 (2008).
19. Li, M., Janner, D., Yao, J. & Pruneri, V. Arbitrary-order all-fiber temporal differentiator based on a fiber Bragg grating: design and experimental demonstration. *Opt. Express* **17**, 19798 (2009).
20. Huang, T. L., Zheng, A. L., Dong, J. J., Gao, D. S. & Zhang, X. L. Terahertz-bandwidth photonic temporal differentiator based on a silicon-on-isolator directional coupler. *Opt. Lett.* **40**, 5614 (2015).
21. Frazier, B. W., Antonsen, T. M., Anlage, S. M. & Ott, E. Wavefront shaping with a tunable metasurface: Creating cold spots and coherent perfect absorption at arbitrary frequencies. *Phys. Rev. Research* **2**, 043422 (2020).
22. Sleasman, T., Imani, M. F., Gollub, J. N. & Smith, D. R. Microwave Imaging Using a Disordered Cavity with a Dynamically Tunable Impedance Surface. *Phys. Rev. Applied* **6**, 054019 (2016).
23. Sleasman, T. *et al.* Implementation and characterization of a two-dimensional printed circuit dynamic metasurface aperture for computational microwave imaging. *IEEE Trans. Antennas Propag.* **69**, 2151 (2020).
24. Kottos, T. & Smilansky, U. Chaotic Scattering on Graphs. *Phys. Rev. Lett.* **85**, 968–971 (2000).
25. Pluhař, Z. & Weidenmüller, H. A. Universal Chaotic Scattering on Quantum Graphs. *Phys. Rev. Lett.* **110**, 034101 (2013).
26. Chen, L., Kottos, T. & Anlage, S. M. Perfect absorption in complex scattering systems with or without hidden symmetries. *Nat. Commun.* **11**, 5826 (2020).
27. Peatman, W. C. B., Wood, P. A. D., Porterfield, D., Crowe, T. W. & Rooks, M. J. Quarter-micrometer GaAs Schottky barrier diode with high video responsivity at 118 μm. *Appl. Phys. Lett.* **61**, 294–296 (1992).
28. Barker, A. S., Verleur, H. W. & Guggenheim, H. J. Infrared Optical Properties of Vanadium Dioxide Above and Below the Transition Temperature. *Phys. Rev. Lett.* **17**, 1286–1289 (1966).
29. Ghanekar, A., Ji, J. & Zheng, Y. High-rectification near-field thermal diode using phase change periodic nanostructure. *Appl. Phys. Lett.* **109**, 123106 (2016).
30. Li, L. *et al.* Electromagnetic reprogrammable coding-metasurface holograms. *Nat. Commun.* **8**, 197 (2017).
31. Stolyarov, A. M. *et al.* Fabrication and characterization of fibers with built-in liquid crystal channels and electrodes for transverse incident-light modulation. *Appl. Phys. Lett.* **101**, 011108 (2012).
32. Gholipour, B., Zhang, J., MacDonald, K. F., Hewak, D. W. & Zheludev, N. I. An All-Optical, Non-volatile, Bidirectional, Phase-Change Meta-Switch. *Adv. Mater.* **25**, 3050–3054 (2013).
33. Wang, D. *et al.* Switchable Ultrathin Quarter-wave Plate in Terahertz Using Active Phase-change Metasurface. *Sci. Rep.* **5**, 15020 (2015).
34. Wang, Q. *et al.* Optically reconfigurable metasurfaces and photonic devices based on phase change materials. *Nat. Photon.* **10**, 60–65 (2016).
35. Ju, L. *et al.* Graphene plasmonics for tunable terahertz metamaterials. *Nat. Nanotechnol.* **6**,

630–634 (2011).

36. Yao, Y. *et al.* Broad Electrical Tuning of Graphene-Loaded Plasmonic Antennas. *Nano Lett.* **13**, 1257–1264 (2013).
37. Huang, Y.-W. *et al.* Gate-Tunable Conducting Oxide Metasurfaces. *Nano Lett.* **16**, 5319–5325 (2016).
38. Bruck, R. *et al.* All-optical spatial light modulator for reconfigurable silicon photonic circuits. *Optica* **3**, 396 (2016).
39. Bello-Jiménez, M. *et al.* Actively mode-locked all-fiber laser by 5 MHz transmittance modulation of an acousto-optic tunable bandpass filter. *Laser Phys. Lett.* **15**, 085113 (2018).
40. Matthès, M. W., del Hougne, P., de Rosny, J., Lerosey, G. & Popoff, S. M. Optical complex media as universal reconfigurable linear operators. *Optica* **6**, 465 (2019).
41. Ou, J. Y., Plum, E., Jiang, L. & Zheludev, N. I. Reconfigurable Photonic Metamaterials. *Nano Lett.* **11**, 2142–2144 (2011).
42. Ou, J.-Y., Plum, E., Zhang, J. & Zheludev, N. I. An electromechanically reconfigurable plasmonic metamaterial operating in the near-infrared. *Nat. Nanotechnol.* **8**, 252–255 (2013).
43. Resisi, S., Viernik, Y., Popoff, S. M. & Bromberg, Y. Wavefront shaping in multimode fibers by transmission matrix engineering. *APL Photonics* **5**, 036103 (2020).
44. Chen, X. *et al.* Active acoustic metamaterials with tunable effective mass density by gradient magnetic fields. *Appl. Phys. Lett.* **105**, 071913 (2014).
45. Xiao, S., Ma, G., Li, Y., Yang, Z. & Sheng, P. Active control of membrane-type acoustic metamaterial by electric field. *Appl. Phys. Lett.* **106**, 091904 (2015).
46. Chen, Z. *et al.* A tunable acoustic metamaterial with double-negativity driven by electromagnets. *Sci. Rep.* **6**, 30254 (2016).
47. Ma, G., Fan, X., Sheng, P. & Fink, M. Shaping reverberating sound fields with an actively tunable metasurface. *Proc. Natl. Acad. Sci. USA* **115**, 6638–6643 (2018).
48. Ao, W., Ding, J., Fan, L. & Zhang, S. A robust actively-tunable perfect sound absorber. *Appl. Phys. Lett.* **115**, 193506 (2019).
49. Tian, Z. *et al.* Programmable Acoustic Metasurfaces. *Adv. Funct. Mater.* **29**, 1808489 (2019).
50. Cao, W. K. *et al.* Tunable Acoustic Metasurface for Three-Dimensional Wave Manipulations. *Phys. Rev. Applied* **15**, 024026 (2021).
51. Zhang, C. *et al.* A reconfigurable active acoustic metalens. *Appl. Phys. Lett.* **118**, 133502 (2021).
52. Zhao, S.-D., Chen, A.-L., Wang, Y.-S. & Zhang, C. Continuously Tunable Acoustic Metasurface for Transmitted Wavefront Modulation. *Phys. Rev. Applied* **10**, 054066 (2018).
53. Fan, S.-W. *et al.* Tunable Broadband Reflective Acoustic Metasurface. *Phys. Rev. Applied* **11**, 044038 (2019).
54. Ma, Z. *et al.* Spatial ultrasound modulation by digitally controlling microbubble arrays. *Nat. Commun.* **11**, 4537 (2020).